**Article**

# Vertex pinning and stretching of single molecule DNA in a linear polymer solution

Kunlin Ma[a], Caleb J. Samuel[a], Soumyadeep Paul[a], Fereshteh L. Memarian[b], Gabrielle Vukasin[b], Armin Darvish[b], Juan G. Santiago[a]*

[a] Department of Mechanical Engineering, Stanford University, Palo Alto, CA 94305.

[b] Robert Bosch LLC, Research & Technology Center, Sunnyvale, CA 94085.

* Corresponding author: Juan G. Santiago

**Email:** juans@stanford.edu

**Abstract**

Trapping, linearization, and imaging of single molecule DNA is of broad interest to both biophysicists who study polymer physics and engineers who build nucleic acid analyzing methods such as optical mapping. In this study, single DNA molecules in a neutral linear polymer solution were driven with an axial electric field through microchannels and their dynamics were studied using fluorescence microscopy. We observed that above a threshold electric field, individual DNA molecules become pinned to the channel walls at a vertex on each molecule and are stretched in the direction opposite to the electric field. Upon removal of the electric field, pinned DNA molecules undergo relaxation within a few seconds to a Brownian coil around the vertex. After 10's of seconds, DNA is released and free to electromigrate. The method enables high quality imaging of single-molecule DNA with high throughput using simple-to-fabricate fluidic structures. We analyze the conditions needed for trapping, relaxation dynamics, and the repeatability of vertex pinning. We hypothesize DNA entangles with neutral linear polymers adsorbed to walls. We hypothesize that a sufficiently high electric force on the DNA is required to expel a hydration layer between the DNA and the wall-adsorbed neutral linear polymers. The elimination of the hydration layer may increase the friction between charged DNA and the uncharged polymer, promoting vertex pinning of DNA.

**Main Text**

Single-molecule DNA trapping and stretching enables fundamental studies of DNA polymer physics and DNA interaction with flow and electric fields[1]. Methods to trap and stretch single-molecule DNA with simultaneous visualization include using optical or magnetic tweezers to manipulate microbeads tethered to DNA[2–4]; subjecting DNA to extensional flow strain in microfluidic devices[5,6]; incorporating linking chemistries to tether DNA to flat surfaces and then subjecting them to fluid flow or electric fields, as in so-called DNA curtains[7,8]; and so-called combing of DNA on silanized glass using a moving air/liquid interface[9,10]. Confining and trapping of single-molecule DNA combined with high quality imaging also has important applications in genetic analyses, including karyotyping and optical DNA mapping[11]. For example, electromigration of DNA with sequence-specific labels within complex nanochannels arrays is used to elongate and image long (>100 kb) genomic DNA for medical diagnostics[12].

We here present a new method that can be used to trap, stretch, release, and re-trap many single-molecule DNA in a simple-to-fabricate device with no moving parts. Trapping of single DNA molecules is achieved using entanglement onto a linear polymer scaffold. A linear polymer is used to create a scaffold on the channel surfaces onto which each DNA is pinned at a vertex along its length. The DNA molecules are trapped, held in place, and stretched by the applied axial electric field. This enables high-quality imaging and quantification of DNA conformation. We here study the necessary conditions for DNA to be trapped; quantify trapping dynamics; and quantify relaxation dynamics, disentanglement, and release from pinned state.

**DNA vertex pin to polyvinylpyrrolidone at sufficiently high electric field**

Figure 1 summarizes DNA trapping phenomena. The channel is etched in silicon, passivated with thermal oxide, and has a 0.9 μm depth and a nominal spanwise width of 20 μm (see Fig. S1 of the Supporting Information, SI, for more details). The channel in Figure 1 includes a castellation pattern on one side, but this feature is not required for the trapping. The trapping also occurs similarly in channels within which all walls are flat and with surface materials of thermally grown oxide, borosilicate and fused silica (see Extended Data Fig. 1 and Extended Data Fig. 2). Figure 1A shows schematics of each DNA pinned at the wall at a vertex with two loose arms stretching in the direction opposite to the applied electric field, consistent with the electrostatic force on the polymer and in the direction opposing any residual electroosmotic flow (EOF). Figure 1B shows consecutive false-color epifluorescence images of single-molecule 48.5 kbp λ-DNA at electric fields of 350, 20 and 0 V/cm. First, DNA traps at high field. Upon lowering the field, the DNA trapping persists and Brownian motion results in a state where DNA arms overlap to a lesser degree. Electric field was deactivated at $t$ = 3.5 s, and DNA molecules then quickly coil into a three-dimensional (3D) cloud centered around the aforementioned vertex pin point (e.g., $t$ = 7 s). The leftmost panel of Figure 1C shows a bright-field image of the channel geometry. The figure also shows three consecutive images that capture the accumulation of vertex-pinned, single-molecule DNA. As we shall discuss, a capture and release process can by cycled, enabling high-throughput capture and visualizations of many DNA in the same field of view. The last two images of Figure 1C show two example subsequent realizations of the trapping process. Similar data for 20 kbp DNA is provided in the SI (c.f. Extended Data Fig. 3).

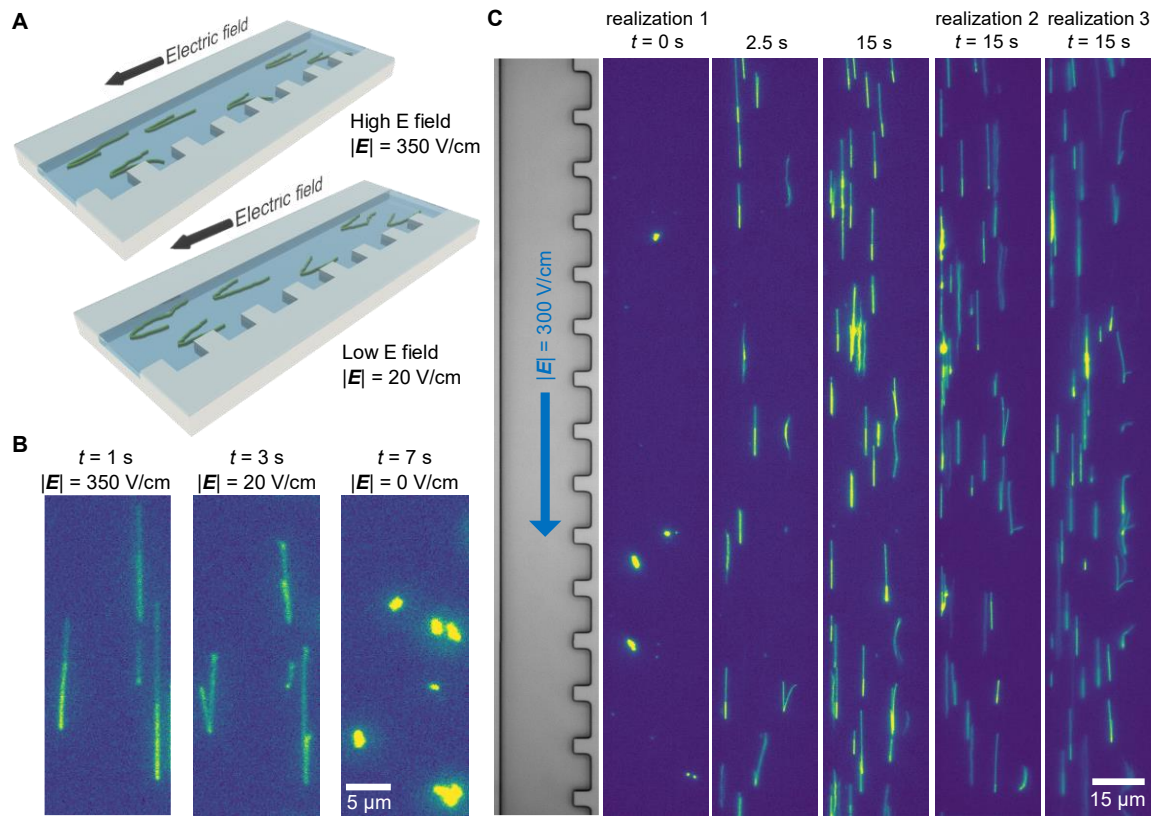

**Fig. 1 | Vertex-pinned, single-molecule 48 kbp DNA in a 0.9 μm deep microfluidic channel filled with a linear polymer buffer solution and subject to an applied axial electric field.** (**A**) Isometric schematics of DNA pinned at a vertex with two arms extending opposite to the applied field. (**B**) Sequential, experimental epifluorescence images of vertex-pinned single-molecule DNA. Images show stretching at 350 V/cm, 20 V/cm, and no applied electric field where DNA relaxes to a Brownian coil around the vertex. (**C**) Three consecutive images of single-molecule DNA pinning upon application of 300 V/cm. The fourth and fifth images are two additional realizations of the trapping process. Solution was 4 pM of YOYO-1 labeled *λ*-DNA in 1×TBE buffer with 2% w/w 1300 kDa PVP and 4% v/v *β*-mercaptoethanol.

We performed experiments at electric fields ranging from 19 to 455 V/cm and at a variety of buffer chemistries, including with and without linear polymer additives. Solutions of neutral, water soluble polymers are typically used to suppress electroosmotic flow. Here we used polyvinylpyrrolidone (PVP) at various molecule weights[13]. We observed DNA trapping via vertex pinning only at sufficiently high electric fields and only in the presence of polyvinylpyrrolidone (PVP) with a molecular weight (MW) of 360 kDa or larger. PVP is used as a dynamic coating which at least partially adsorb to glass and oxide channel walls, thereby increasing the bulk velocity of the solution within the electric double layer[14,15]. We hypothesize that adsorbed PVP molecules present a scaffold whereupon individual DNA molecules become entangled. This entanglement most often occurs at a single point along the DNA. The pinning occurs on the silicon oxide wall of the channels and on the surface of the glass wall used to seal the channels. The latter is supported by observations we made in channels of various depths wherein the imaging focal plane was placed at various distances from the top and bottom walls (see Extended Data Fig. 1). Figure 2 shows a schematic of this hypothesized mechanism. PVP polymers are adsorbed to both the top wall and bottom wall. DNA molecules (normally in a Brownian coil) are driven by electric fields. These DNA can frequently interact with and hook onto the adsorbed polymer, especially in channels with sufficient confinement. The entanglement most often results in pinning at a single vertex. For the images in Fig. 1, the depth of field of our optics is about 0.7 µm while the channel transverse depth is 0.9 µm, so all DNA molecules are in focus.

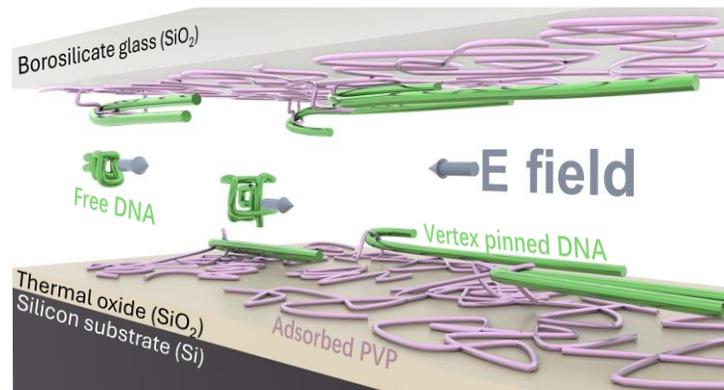

**Fig. 2 | 3D schematic drawing of hypothesized DNA entanglement with wall adsorbed PVP polymer and entanglement at a vertex and subject to an electric field parallel to the wetted surfaces.** The bottom surface is silicone oxide thermally grown on silicon wafer. The top surface is borosilicate glass. The larger green tubes represent DNA molecules, and the adsorbed pink tubes represent PVP polymers. Free DNA molecules travel against the direction of electric fields and some DNA molecules become vertex-pinned to the PVP polymers that are adsorbed to the oxide walls.

We explored the effects of applied electric field magnitude, $E$, on the initiation and rate of DNA trapping. We developed and used a custom image processing algorithm to quantify the number of DNA pinned as a function of $E$. Figure 3A summarizes this analysis and additional details are provided in Section S1.1 and S1.2 of the SI. Briefly, we computed moving median images from small groups of images to enhance signal from stationary DNA (three-image sequences for the largest $E$ and 21-image sequences for the lowest $E$). We then apply adaptive, local thresholding to obtain a binarized image. The binarized image was used as the input for an alpha-shape identification algorithm followed by dilation and erosion to obtain an alpha shape bounding mask[16]. This mask was applied to the aforementioned moving median image sequences and the mask/image product was spatially integrated to obtain a scalar measure (versus time) of stationary DNA for each $E$. In Figure S2, we show this scalar correlates well with example manual counts of trapped DNA (obtained by analyzing each frame manually). We computed the median of the scalar time-series data from our automated analysis over 30 s durations for each $E$. Figure 3B shows these median scalar measures as a function of $E$ for 4 pM of YOYO-1 labeled 48.5 kbp λ-DNA in 1×TBE buffer with 4% v/v β-mercaptoethanol. Shown are data for 2% w/w PVP with MW of 1300 (orange circles) or 360 kDa (green squares). Each of the $E$ field data (for 360 and 1300 kDa) exhibited an approximate threshold $E$ value required to initiate trapping (see Extended Data Fig. 4 for $E$ field normalized version of Fig. 3B and Fig. S3 in the SI for the time evolution data). The observed thresholds were 70 V/cm for the 1300 kDa PVP and 180 V/cm for the 360 kDa PVP (see Movie S1 and Movie S2 for additional details). The tension force $T$ acting on DNA due to electric field can be estimated as $T = E\,\rho_{\text{eff}}\,h$[17]. Here, $E$ is the field strength, $h$ is the projection of an arm length along the direction opposite to the electric field, and $\rho_{\text{eff}}$ is the electrophoretic charge density of the DNA molecule. The electrophoretic charge density, $\rho_{\text{eff}}$, is on the order of 0.05 e⁻/phosphate in free solution (where e⁻ is an elementary charge)[17]. For our threshold field strength of 70 V/cm field and 48.5 kbp DNA in 2% w/w 1300 kDa PVP, we estimate the tension force exerting on DNA to be 1.4 pN assuming $h$ is approximately half of the DNA contour length. This tension force range is much lower than the reported 35 pN tension force required to break individual λ-concatemers by direct pulling or the reported 15 pN to nick and create so-called *cos* sites[17,18]. The PVP polymers adsorb to oxide walls mostly due to hydrogen bonds, and the adsorption forces are expected to be much larger than the tension forces exerting on DNA molecules by the electric field[19–21]. Larger electric fields result in a higher trapping rate. We note that we also performed experiments using buffers with 2% w/w PVP with MWs of 10 and 58 kDa PVP. We observed no DNA trapping for solutions prepared using these lower, commonly used, commercially available PVP molecular weights. In addition, we observed no trapping under any pressure driven flow in the absence of electric fields. Referring again to the aforementioned PVP entanglement hypothesis, these observations suggest that sufficiently long PVP is required for PVP to adsorb to the surface of the channel and present a PVP polymer scaffold that enables DNA entanglement onto PVP. Furthermore, we hypothesize that trapping rate and also the fraction of DNA trapped will depend on the concentration of PVP and the channel transverse depth. More PVP added to the buffer may provide more trapping sites. Further, the degree of DNA confinement should affect the rate at which DNA interacts with the wall. Observations of DNA trapping in channels with less confinement (e.g., 3 µm depths as in Extended Data Fig. 1) confirm that a lesser fraction of DNA entering field of view are trapped.

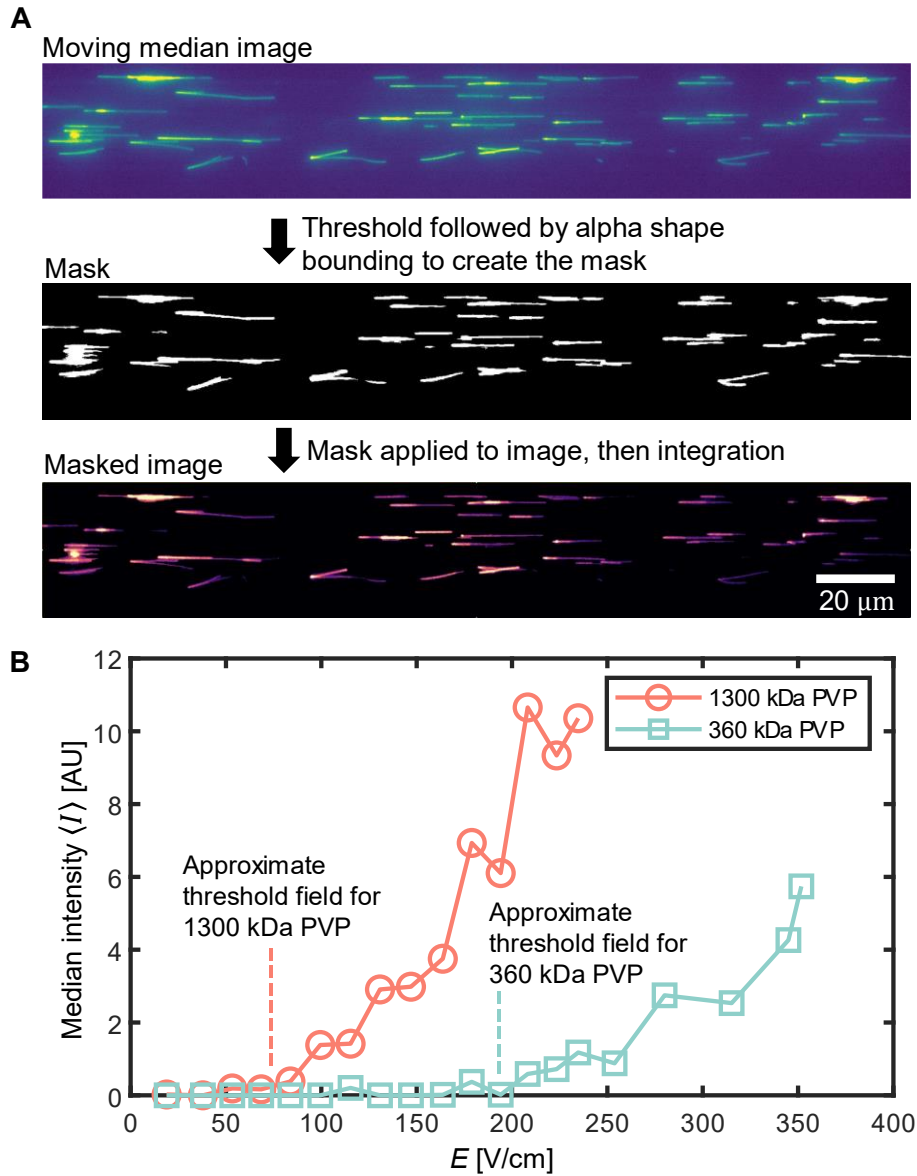

**Fig. 3 | Experimental quantification of the amount of vertex-pinned, single-molecule DNA in a 0.9 µm deep channel as a function of field strength.** (**A**) A summary of image processing used to quantify the amount of DNA trapping. Medians of small numbers of images were subsequently adaptively thresholded. The binarized images were used to create alpha shape bounding masks, and raw image data integrated within the regions highlighted by binary image masks. (**B**) Area-averaged image intensity, ⟨$I$⟩, for 30 s duration experiments at each applied E. All data are for 4 pM of YOYO-1 labeled $\lambda$-DNA (48.5 kobo) in 1×TBE buffer with 4% v/v $\beta$-mercaptoethanol. The orange and green curve show experiments using solutions prepared with

PVP MWs of MWs of 1300 or 360 kDa, respectively (both 2% w/w concentration). Dashed vertical lines highlight approximate threshold fields observed for the initiation of DNA trapping.

The data suggests that a high magnitude electric field is required for high friction force between the DNA and PVP polymers. Given the threshold nature of the required field, we hypothesize that a sufficiently high electric force on the DNA is required to expel a hydration layer between the DNA and PVP molecules. Elimination of a hydration layer may increase the solid friction between charged DNA polymer and the uncharged linear PVP polymer[22,23]. This hypothesis is consistent with the fact that persistent pinning can occur at a vertex near the end of a DNA molecule such that the two loose arms have widely disparate lengths and should experience much different electrostatic forces from the electric field. The elimination of hydration layers between single DNA molecules and polymer molecules has been hypothesized in studies of DNA interacting with agarose cross-linked polymer[17,24–26]. For example, in studies of DNA electrophoresis through 3D agarose gel network, Bustamante and Volk concluded that DNA electromigration can be strongly hindered and even arrested under high electric fields [17]. Similarly, Viovy et al. observed trapping of DNA within an agarose gel for electric fields above certain threshold of electric fields and proposed the "knot" trapping hypothesis [26]. However, the entangled DNA in these previously studies acquired a complex 3D shape within the interior of the gel and not a "clean" vertex pinning at a wall with two straight arms (within a linear polymer solution) as in the current work. Further, the actual mechanism of such trappings is still not yet fully resolved.

**Relaxation of vertex-pinned DNA**

We analyzed the morphology, position, and relaxation dynamics of vertex-pinned DNA upon removal of the electric field. Figure 4A shows an example time-sequence of images. We observed DNA pinned at random locations along the length of the DNA molecule, and this led to a range of relaxation timescales. To analyze these, we developed an automated image processing method based on adaptive thresholding, fitting of DNA images with an ellipse, and extracting of the major and minor axes of the best-fit ellipses (see discussion in the SI). Figure 4B shows a close-up image with best-fit ellipses.

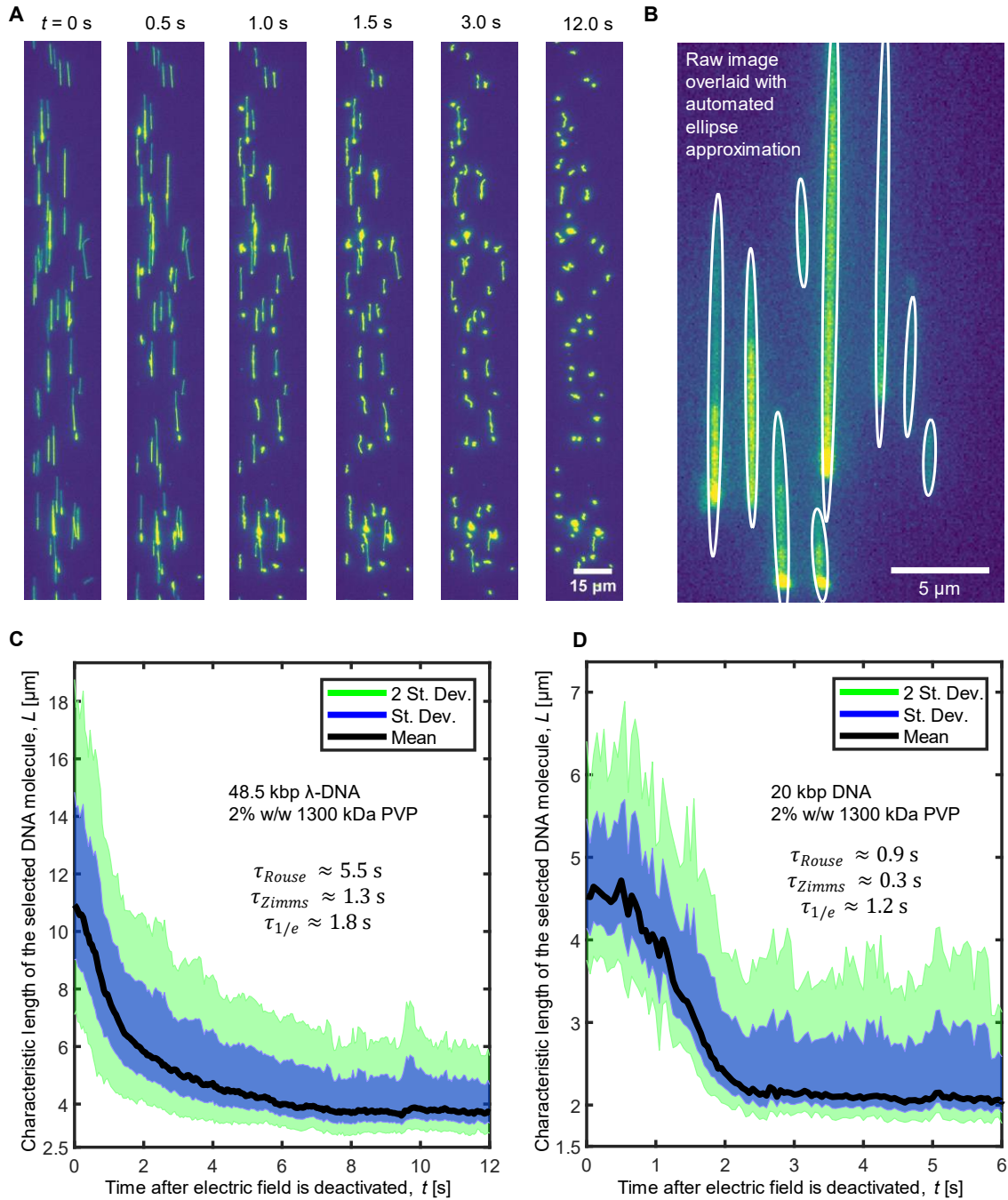

**Fig. 4 | Relaxation dynamics of ensembles of vertex-pinned single-molecule DNA.** (A) Example sequence of raw epifluorescence microscopy images of pinned λ-DNA (48.5 kbp) immediately after deactivation of the applied electric field (at $t = 0$ s). (B) Example raw image with superposed best-fit ellipses used to quantify DNA length and width. (C) and (D) show plots of the statistics of ellipse major axis lengths, $L$, for relaxation of 48.5 and 20 kbp DNA molecules versus

the time since deactivation of the field. Black curve shows the instantaneous mean of *L*, and blue and green shaded curves show one and two standard deviations of the distributions, respectively. The Rouse relaxation timescale and Zimms relaxation timescale for fully stretched 48.5 kbp DNA are 5.5 and 1.3 s, respectively. The Rouse relaxation timescale, $\tau_{Rouse}$, and Zimms relaxation timescale, $\tau_{Zimms}$, for fully stretched 20 kbp DNA are 0.9 and 0.3 s, respectively. By comparison, the characteristic 1/e timescale for relaxation observed for our ensembles of 48.5 and 20 kbp DNA, $\tau_{1/e}$, are approximately 1.8 and 1.2 s, respectively.

Figure 4C and 4D show statistical moments of characteristic lengths of pinned DNA molecules as they relax (upon removal of the electric field). Figure 4C and 4D show data for 48.5 or 20 kbp DNA, respectively, and both labeled with YOYO-1 dye in 1×TBE buffer of 4% v/v β-mercaptoethanol and 2% w/w PVP of MWs of 1300 kDa. The initial mean values for *L* were lower than the expected full contour lengths as most DNA were pinned at a vertex well away from the end of the molecule. The observed lengths for the nominal 48.5 kbp DNA immediately decays from the initial value (c.f. Fig. 4C). In contrast, the shorter 20 kbp length data (Fig. 4D) show a concave down inflection before decaying rapidly. The latter is reproducible behavior observed only for the 20 kpb data. Both sets of data asymptote to a steady state consistent with a Brownian coil around the vertex pin point. The characteristic 1/e timescale for the observed relaxation dynamics for our ensembles of 48.5 and 20 kbp DNA are $\tau_{1/e}$ = 1.8 and 1.2 s, respectively. To provide a comparison, we calculated two relaxation timescales based on two classic polymer physics theories: Rouse and Zimms relaxation timescales. The Rouse relaxation timescale is given as $\tau_{\text{Rouse}} = \frac{\eta_s b^3}{k_B T} N^2$ and the Zimms relaxation timescale is given as $\tau_{\text{Zimms}} = \frac{\eta_s b^3}{k_B T} N^{3\nu}$. Here, $\eta_s$ is the kinematic viscosity of the solvents, $b$ is the persistence length, $k_B$ is the Boltzmann constant, $T$ is the temperature, $N$ is the number of segments, and $\nu$ is the Flory exponent[27,28]. We chose $\nu = 0.588$ to reflect the current good solvent situation (water). According to the classic theory of polymer chain relaxation, for fully stretched 48.5 kbp DNA, the Rouse and Zimms relaxation timescales are $\tau_{\text{Rouse}} \approx 5.5$ s and $\tau_{\text{Zimms}} \approx 1.3$ s, respectively. For fully stretched 20 kbp DNA, the Rouse and Zimms relaxation timescales are $\tau_{\text{Rouse}} \approx 0.9$ s and $\tau_{\text{Zimms}} \approx 0.3$ s, respectively. Epi-fluorescence images of relaxation of vertex-pinned 20 kbp DNA are included in Extended Data Fig. 5. The movies of relaxation of vertex-pinned 48.5 and 20 kbp DNA are included in the SI (Movie S3 and S4). As described above, upon removal of the (high) electric fields, the DNA consistently relaxes toward and around the vertex of the pinned location. This further supports our hypothesis of a single-point entanglement between DNA and PVP. The persistence of such a pinning point (for order of 30 s) suggests that the observed DNA trapping is not due to electrosorption forces such as dielectrophoretic forces which might manifest from wall imperfections (e.g., roughness elements). Such electrosorbed DNA molecules would be expected to quickly "release" upon deactivation of the field. Note that DNA were not trapped at the convex corners of castellation features in the channel. During the first 2 s of relaxation, the shape of the 20 kbp DNA relaxation curves is qualitatively different than that of 48.5 kbp DNA, and this is repeatable across initial electric fields and realizations. We hypothesize that this may be due to the larger average electrostatic forces experienced by the 48.5 kbp DNA. The longer DNA arms may result in a higher degree of elastic stretching in the initial, stretched state.

**DNA vertex-pinning, relaxation, and disentanglement cycle**

We demonstrate cycles of the DNA process across batches of DNA molecules. The cycle includes vertex-pinning at high electric field; relaxation upon deactivation of field; disentanglement (here after about 50 s); and then a state of free diffusion and electromigration of DNA. The latter state can be leveraged to remove previously pinned DNA and introduce new DNA for the next cycle. In a typical cycle c.f. Extended Data Fig. 6), DNA molecules are initially in a three-dimensional Brownian coil. Then a high electric field (*E* = 190 V/cm in Extended Data Fig. 6) is activated to initiate trapping. After applying high electric field for roughly 30 s for trapping, the electric field is deactivated and DNA molecules quickly relaxed toward their vertex pinpoint. Over a 50 s time scale, relaxed DNA become unpinned from the wall and were again free to diffuse and electromigrate. Lastly, a low axial electric field of 19 V/cm was applied to clear the field of view (FOV) of previously pinned DNA (and introduce new DNA). The process was repeated in multiple, subsequent cycles with an estimated period of order 90 s. See Movie S5 for an example of DNA vertex-pinning, relaxation and disentanglement cycle.

The data of Extended Data Fig. 6 further supports the hypothesized entanglement between DNA and PVP polymer. After removal of the electric field, we hypothesize the space between the DNA and PVP becomes re-hydrated, and the associated lower-friction configuration and Brownian motion promote a disentanglement after several 10's of seconds. We also observed that a pulse of pressure-driven flow can cause much faster disentanglement down to order of a few seconds (data not shown).

Lastly, we studied whether DNA pinning sites are spatially correlated across multiple realizations within the same field of view. Figures 5A-5C show three representative realizations with three respective false-color pallets (Realizations 1, 2, and 3). Between realizations, DNA were allowed to fully relax and disentangle and were cleared with low electric fields. Figure 5D shows superposition of Realization 1 and 2 images; and Figure 5E shows superposition of Realization 1, 2 and 3. Zoomed-in images of Figs. 5D and 5E are shown in Figure 5F and 5G.

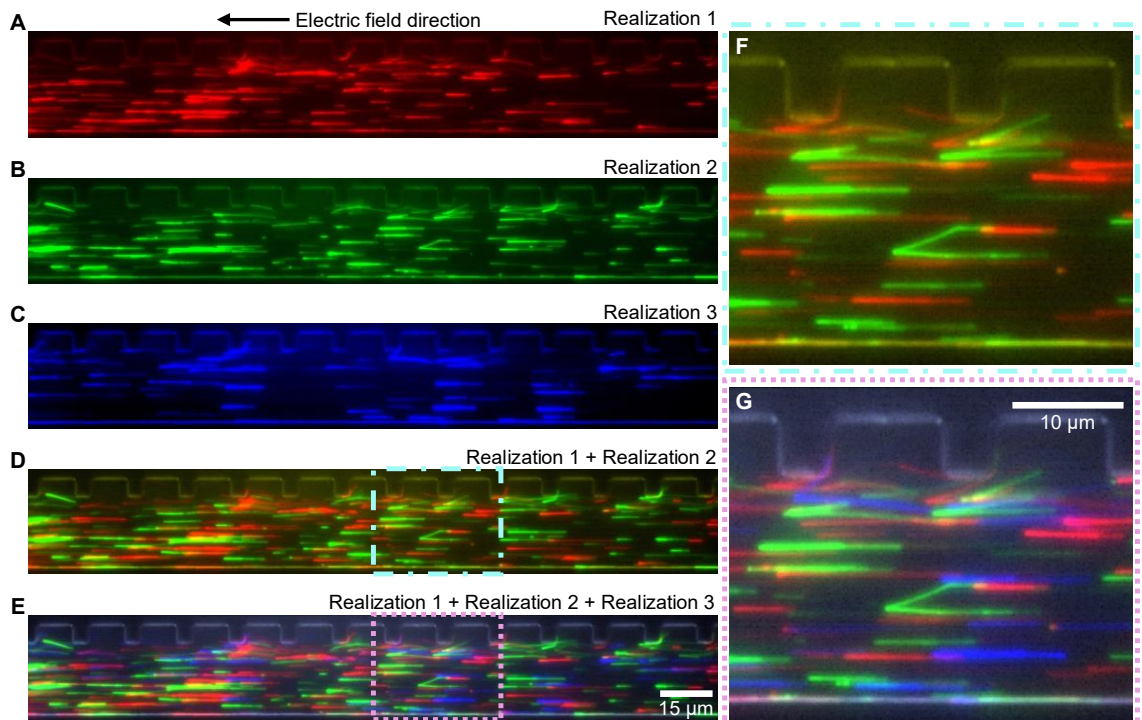

**Fig. 5 | Single DNA molecules are vertex-pinned to walls at locations which vary randomly from realization to realization. All data are 4 pM of YOYO-1 labeled λ-DNA with 2% w/w 1300 kDa PVP.** (**A** to **C**) Images of three separate realizations of DNA trapping in the same field of view using red, green, and blue false-color maps. DNA were trapped at 300 V/cm for 15 s. DNA were allowed to relax and disentangle between experiments, and a low electric field was used to both confirm disentanglement and to clear previously trapped DNA. (**D** and **E**) Superposed false-color images of Realizations 1, 2 and 3 showing poor spatial correlation of vertex capture points across two and three realizations, respectively. (**F**) and (**G**) are zoomed-in images of (**D**) and (**E**), respectively.

We observed a poor correlation of DNA capture locations across all realizations studied. Statistical colocalization analysis of DNA trapping across realizations were performed, and the results are shown in Fig. 6 (details in Section S1.3 of the SI). The Pearson coefficients are defined as $R = \frac{\sum (R_i) \cdot (G_i)}{\sqrt{\sum (R_i)^2 \cdot \sum (G_i)^2}}$, in which $R_i$ are the intensities of the first color on each pixel of index $i$ and $G_i$ are the corresponding intensities for the second color on the same pixel of index $i$[29,30]. The calculated Pearson coefficients are 0.30, 0.25 and 0.22 among three independent realizations from Figure 4A, 4B, and 4C. Compare these Pearson coefficient values to those of highly correlated images obtained at various times within a single realization (see Section S1.3 of SI). These results and analysis further support our hypothesis that the observed trapping is not due to specific features in the channel geometry such as roughness elements or nano-scale surface pits. The data supports our hypothesis that DNA molecules become entangled with adsorbed PVP molecules. This entanglement is eventually reversible and newly introduced DNA molecules become pinned at new, uncorrelated locations. Lastly, we note that the approximately linear growth of the trapping rates shown in Fig. 3 suggests that the trapping sites for DNA do not reach saturation for the duration of our experiments. The latter suggests a relatively large number of available trapping sites.

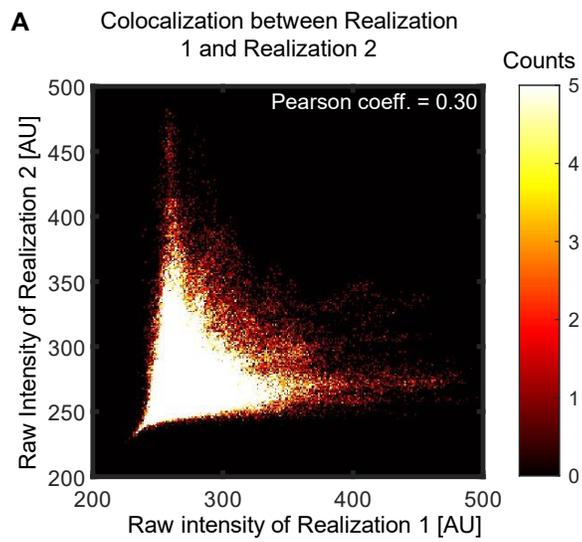

**A** Colocalization between Realization 1 and Realization 2

Pearson coeff. = 0.30

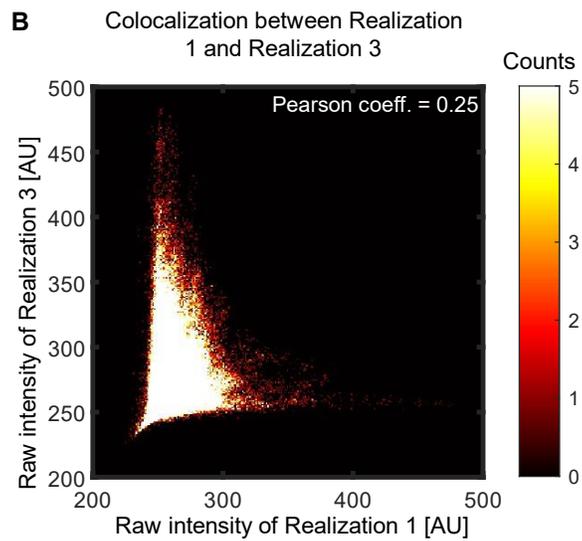

**B** Colocalization between Realization 1 and Realization 3

Pearson coeff. = 0.25

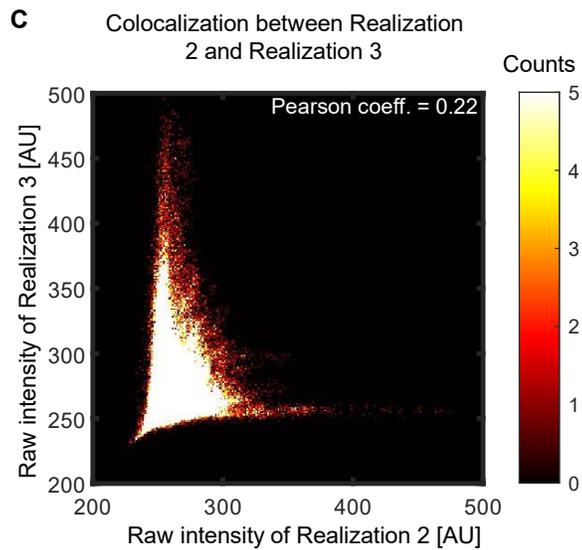

**C** Colocalization between Realization 2 and Realization 3

Pearson coeff. = 0.22

**Fig. 6 | Colocalization analysis of image intensity of vertex-pinned DNA across three different realizations.** Colocalization data are shown for three pairs of independent realizations of DNA trapping. Realizations 1, 2 and 3 are DNA vertex-pinned DNA data from Fig. 5A, 5B and 5C in the main text, respectively. (**A** to **C**) Colocalization scatter plots (correlation diagrams) are shown comparing each pair of realizations. As indicated by the intensity bar on the right, the heat map are the counts based on raw intensity values on the collocated pixels from two images of two realizations. Pearson correlation coefficients are computed between each pair, and the values are 0.30, 0.25 and 0.22, respectively.

**Discussion**

Two necessary conditions for the observed trapping via vertex pinning and stretching of DNA seem to be large PVP MW (here 360 kDa and longer) and sufficiently large applied axial electric field. Observed threshold fields are smaller for large MW PVP. The phenomenon occurs only at the surface and occurs on both thermal silicon oxide and glass surfaces (see Extended Data Fig. 1 and Movie S6). We observed no such trapping (at any electric field) for solutions of PVP of MW of 10 and 58 kDa PVP. We also observed no trapping of any kind upon applied pressure driven flow (no electric field). See Movie S7 for more details. We hypothesize that pressure driven flow may not result in sufficient forces between the DNA and PVP to expel a hydration layer. Also, pressure driven flow may distort and strain the PVP so that entanglement is strongly impeded.

We posed several hypotheses. First that, at sufficiently high electric fields, DNA become entangled with surface-adsorbed PVP polymer, and the strong electric forces expel a hydration layer between DNA and PVP, resulting in a high friction condition. We estimated the threshold tension force exerted on DNA is much lower than forces that reportedly nick or break DNA strands. We also hypothesize this electric force on the DNA is significantly lower than the force between the linear neutral polymer and the oxide walls (via hydrogen bonds). Upon removal of the electric field, the space between DNA and PVP may become re-hydrated, and the resulting lower friction enables Brownian motion to disentangle DNA from PVP. We observed no spatial correlation of vertex pin-points across realizations, suggesting that the phenomenon is not associated with specific channel geometry (e.g., roughness elements).

We demonstrated a cycle of vertex-pinning, relaxation, disentanglement and free electromigration to trap, linearize, and visualize many single-molecule DNA. Our data show that vertex pinning and stretching can be interspersed with relaxation, disentanglement, and low-electric-field clearing of DNA to cycle and obtain high-quality images of many DNA. We suggest that the system can be used as a platform to study various DNA dynamics. Further, it can potentially be engineered and combined with sequence-specific discrete labeling to achieve DNA optical mapping.

## Materials and Methods

**Custom microfluidic interface device.**
   Figure S1A shows the three main components of our custom microfluidic device. These are four PDMS reservoirs, a borosilicate glass slip (channel cover), and a silicon substrate chip. Figure S1B shows an isometric view of the device after bonding and assembly. Figure S1C shows an image of the experiment showing the chip, platinum electrodes in reservoirs (connected to alligator clips), and the water immersion objective. Figure S1D shows the layout of the channels in the silicon substrate chip. The chip has two isolated (independent) single-input, single-output channels. The channels were fabricated so that they could be visualized simultaneously within the same field of view near the geometric center of the silicon chip. Fabrication was carried out at the Stanford Nanofabrication Facility (SNF) on 4-inch n-type silicon wafers (resistivity of 4 Ohm-cm). Following standard photolithography, the microfluidic channels were dry-etched into the silicon wafer to a depth of 1 μm (actual resulted depth was 0.9 μm measured by profilometer) using HBr and $BCl_3$ (Oxford Instruments PlasmaLab III-V etcher). Residual photoresist was removed using oxygen plasma which was followed by a 20-minute piranha cleaning at 120 ℃. 300 nm thick thermal oxide was then grown on the wafers by Rogue Valley Microdevices. The wafers were diced with a wafer saw to 9 devices each with in-plane dimensions of 22 by 24 mm. These devices were then subjected to another round of piranha cleaning. The glass cover slips were No. 1 thickness borosilicate 3.3 glass (purchased from Kemtech America Inc.). The glass drilling was performed using 1.1 mm diameter, triple-ripple diamond drill bits (purchased from Arrowhead Lapidary & Supple) and a manually actuated drill press. The drilled cover slips were anodically bonded as a service provided by Microfluidic Foundry (Berkeley, CA). After the anodic bonding, custom PDMS reservoirs of 5.0 mm OD and 1.5 mm ID were cut using biopsy punches and then bonded to the glass using air plasma. The drilled holes in the glass were aligned with the cut holes on PDMS to form a fluidic connection.

**Sample preparation**
   λ-DNA samples were purchased from ThermoScientific (Waltham, MA) at a stock concentration of 0.3 μg/uL. 20 kbp DNA samples were purchased from ThermoScientific and at a stock concentration of 0.5 μg/uL. YOYO-1 dye was purchased from Invitrogen at a stock concentration of 1 mM. 10×TBE buffer was purchased from Invitrogen and consisted of 1.0 M Tris, 0.9 M Boric acid, and 0.01 M EDTA. Polyvinylpyrrolidone (PVP) of molecular weight (MWs) of 10, 58, 360 and 1300 kDa were purchased from ThermoScientific. We note we know of no commercially available PVP with MWs between 58 and 360 kDa and between 360 kDa and 1300 kDa which we were able to find commercially available. β-mercaptoethanol (MW = 78.13 Da, >98.0% purity) was purchased from TCI America. 10×TBE buffer was initially diluted to 1×TBE buffer with molecular biology grade water (purchased from Fisher Bioreagents, Pittsburgh, PA). Next, β-mercaptoethanol was added to 1×TBE buffer to achieve 2% v/v concentration as an oxygen scavenger. PVP was then added at each aforementioned MW to the buffer at a concentration of 2% w/w. The final resulting buffers consisted of 1×TBE, 4% v/v β-mercaptoethanol, and 2% w/w of PVP with MWs of 10, 58, 360, or 1300 kDa, respectively. We measured a conductivity of 1487 μS/cm of the loading buffers via a conductivity probe (Advanced Electrochemistry Meter from ThermoScientific Inc.). The previous steps were all performed at room temperature (20±1 ℃). Next, raw DNA samples were diluted in the loading buffers to achieve a final concentration of 4 pM, and YOYO-1 dye was added to achieve a base-pair-to-dye stochiometric ratio of 5:1. Samples were then each incubated at 40 ℃ for 1 hour prior to use to facilitate dye intercalation.

**Measurement apparatus**
   Figure S1C shows an example image of one experiment using the assembled microfluidic device. The visualizations were performed using a standard OLYMPUS BX60 upright epifluorescence microscope. We used a water-immersion 60× objective with a numerical aperture of 1.2 (OLYMPUS UPlanApo). The illumination source was a high-power blue LED from ThorLabs (SOLIS-470C) controlled with a ThorLabs DC200 controller, and the epifluorescence filter cube has the following excitation/dichroic/emission components: EX460-490, DM505, EM510IF (OLMPUS U-MWIB2). Two cameras were used for image acquisition. The first camera was a scientific complementary metal-oxide-semiconductor (sCMOS) camera manufactured by Hamamatsu (ORCA-Flash 4.0LT) which we controlled using HCImageLive software. The sCMOS camera was used to acquire all data in the main manuscript and the majority of the data in the Supplementary Information (SI). The second camera was an Andor electron multiplying charge-coupled device (EMCCD) camera (iXon Ultra 897) which we controlled using Andor SOLIS software. The EMCCD camera was used to acquire image data for Extended Data Fig. 2 (images within a wet-etched all glass channel described below).

Voltage was applied using platinum electrodes inserted at the end-channel reservoirs. The electrodes were platinum wire (0.368 mm diameter, hard, 99.95 metals basis) purchased from ThermoScientific. Voltage was sourced and current measured using a Keithley 2400 SourceMeter. The SourceMeter was triggered and controlled via RS232 interface by a personal computer (PC) using custom scripts we wrote in MATLAB (MATLAB 2023b, Mathworks Inc., Natick, MA, USA).

**Data availability**
Source data are provided with this paper. Raw data of movies are available upon request from the corresponding author.

**Code availability**
The codes used for data collection and image processing are available upon request from the corresponding author.

**Acknowledgments**
We acknowledge Dr. Mingqiang Yin for anodic bonding of microfluidic channels. We thank Zethea Inc. for loan of the Hamamatsu sCMOS Orca-Flash 4.0LT camera. We gratefully acknowledge Robert Bosch LLC with Dr. Christoph Lang as program manager.

**Author Contributions**
Conceptualization: J.G.S.; Methodology: J.G.S., K.M., A.D.; Investigation: K.M., S.P., C.J.S., G.V., F.M.; Visualization: K.M.; Writing – original draft: K.M., J.G.S., S.P., C.J.S.; Writing – review & editing: all authors.

**Competing Interest Statement**
J.G.S., K.M., S.P. and A.D. are listed as coinventors on a pending provisional patent application related to this work filed at the U.S. Patent and Trademark Office (no. 63/702,095). Other authors declare no competing interests.

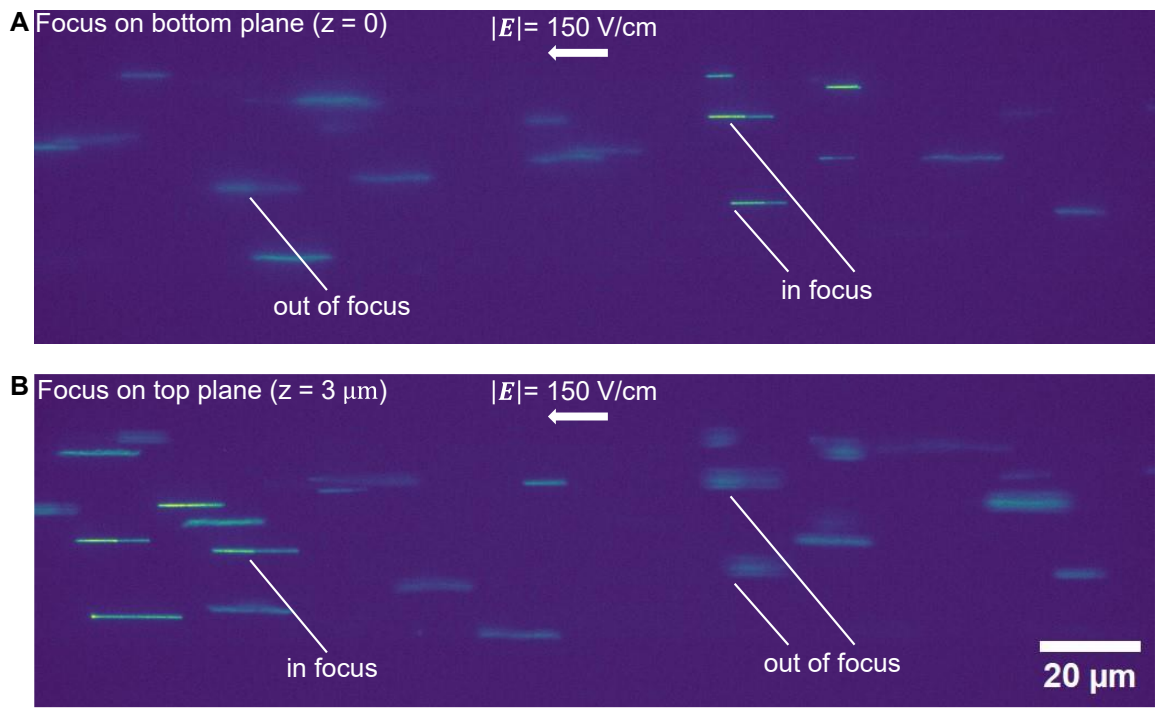

**Extended Data Fig. 1 | DNA vertex-pinning at bottom and top surface of a 3 μm deep channel under axial electric field strength of 150 V/cm.** (**A**) The image shows vertex-pinned DNA at the bottom surface, which is thermally grown thermal oxide. (**B**) The image shows vertex-pinned DNA top surface, which is anodically bonded borosilicate glass. (**A**) and (**B**) are two consecutive images from one realization. The depth of field is estimated to be 0.7 μm.

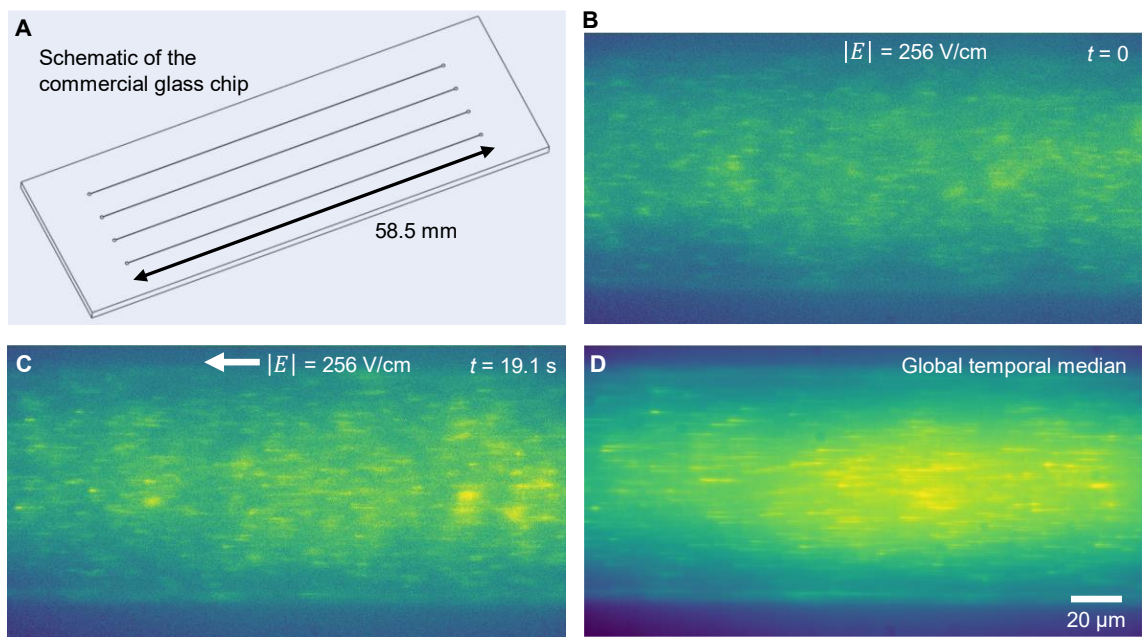

**Extended Data Fig. 2 | Vertex-pinned, single-molecule DNA in a 37 μm deep commercial glass microfluidic channel filled with a linear polymer solution and subject an axial electric field.** (**A**) Schematic of the commercial glass chip purchased from Microfluidic ChipShop. On each chip, four 58.5 mm long parallel glass channels are fabricated and sealed with 210 μm lids. Both substrate and cover are fused silica glass. (**B** to **C**) Consecutive, epifluorescence raw images of DNA electromigrate through 37 μm deep commercial glass channel. The depth of field is estimated to be around 1 μm, and most DNA are out of focus. In the background near to the wall surface, vertex-pinned DNA can be observed. See Movie S10 for more details. (**D**) Global temporal median of the image sequence. The image amplifies the vertex-pinned DNA at the wall. See Movie S10 for more details.

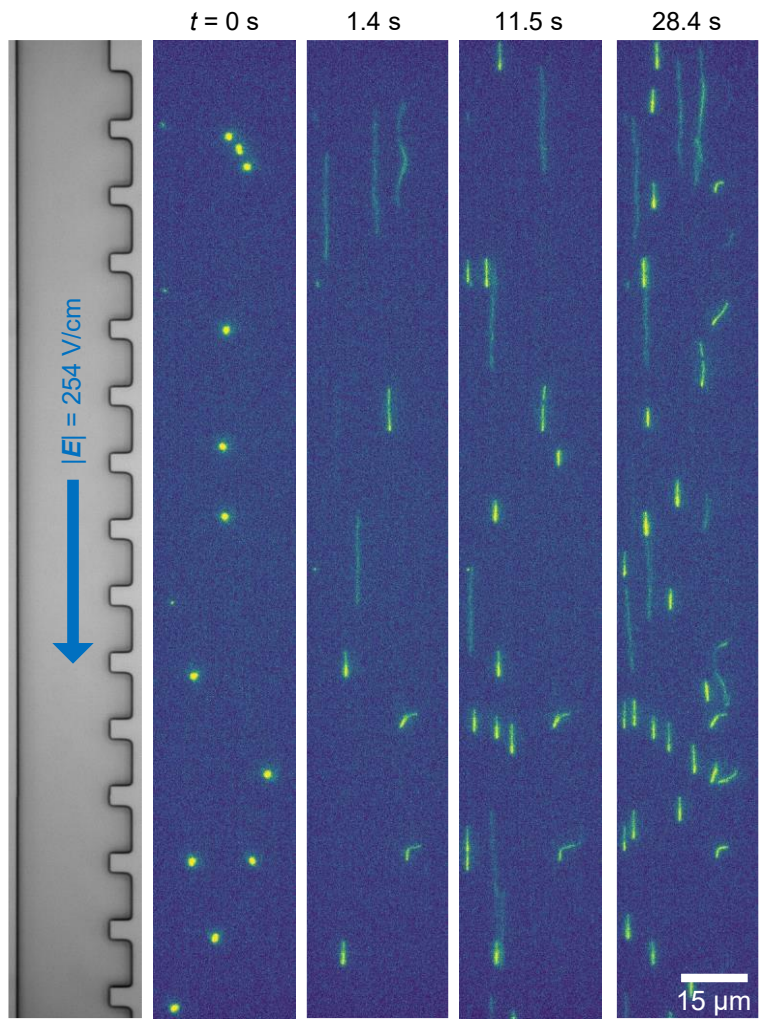

**Extended Data Fig. 3 | Vertex-pinned, single-molecule 20 kbp DNA in a 0.9 µm deep microfluidic channel filled with a linear polymer buffer solution and subject to an applied axial electric field.** Wide-field, consecutive raw images of single-molecule DNA pinning upon application of 254 V/cm. Solution was 4 pM of YOYO-1 labeled 20 kbp DNA in 1×TBE buffer with 2% w/w 1300 kDa PVP and 4% v/v $\beta$-mercaptoethanol. Raw images are shown for $t$ = 0, 1.4, 11.5 and 28.4 s. The faint lines in the background are due to streaking of 20 kbp DNA under fast electromigration.

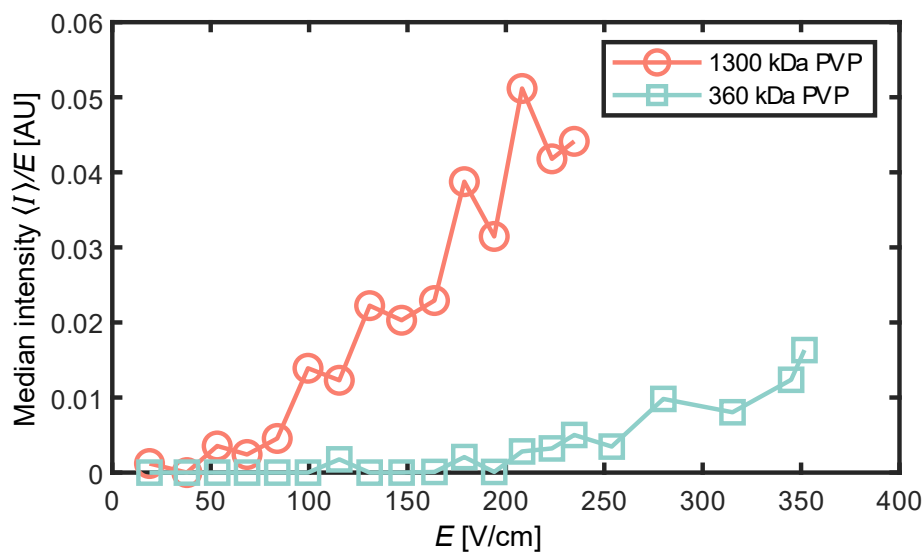

**Extended Data Fig. 4 | Experimental quantification of the amount of vertex-pinned, single-molecule DNA as a function of field strength.** Median intensity of each realization of applying electric field for 30 s are plotted on the ordinate versus the axial electric field strength of each realization on the absicca. These data are comparable to that of Figure 3B in the main text. Here, unlike Figure 3B, the integrated intensity of pinned DNA was normalized by the applied axial field strength. All data are 4 pM of YOYO-1 labeled $\lambda$-DNA (48.5 kbp) in 1×TBE buffer with 4% v/v $\beta$-mercaptoethanol. The orange curve and green curve show experiments of using sample of adding 2% w/w concentration of PVP with MWs of 1300 and 360 kDa, respectively.

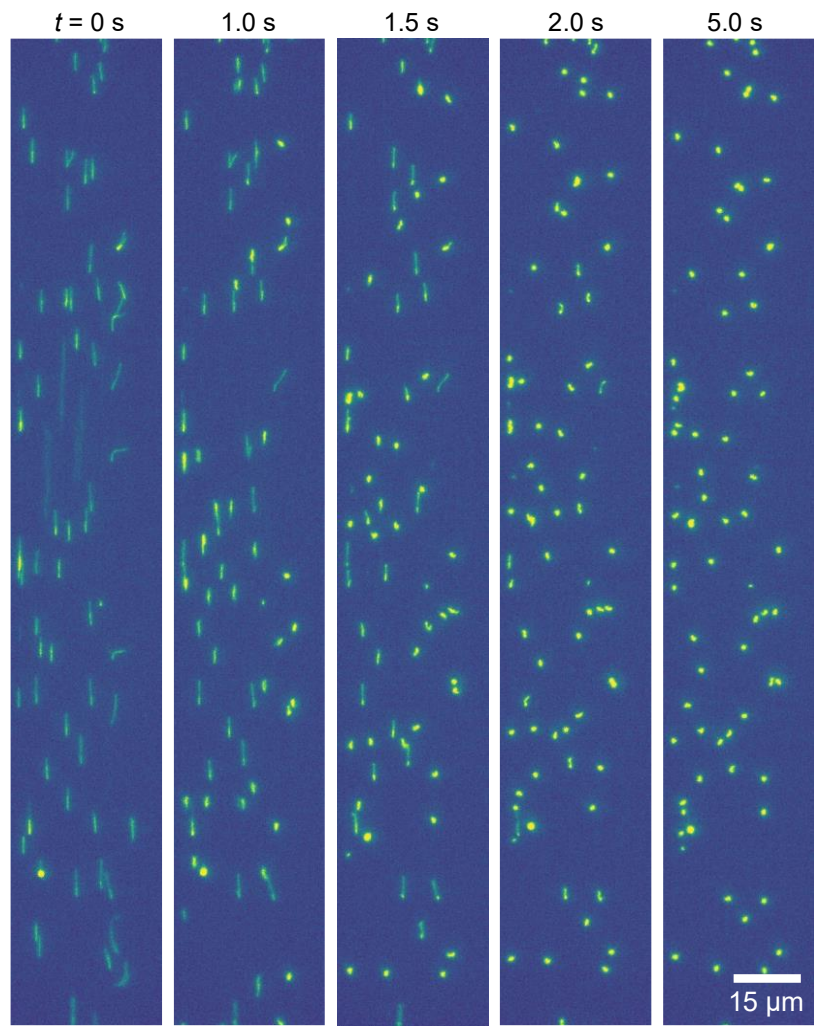

**Extended Data Fig. 5 | Relaxation dynamics of vertex-pinned, single-molecule 20 kbp DNA.**
Example sequence of raw epifluorescence microscopy images of pinned DNA immediately after deactivation of the applied electric field (at $t = 0$ s).

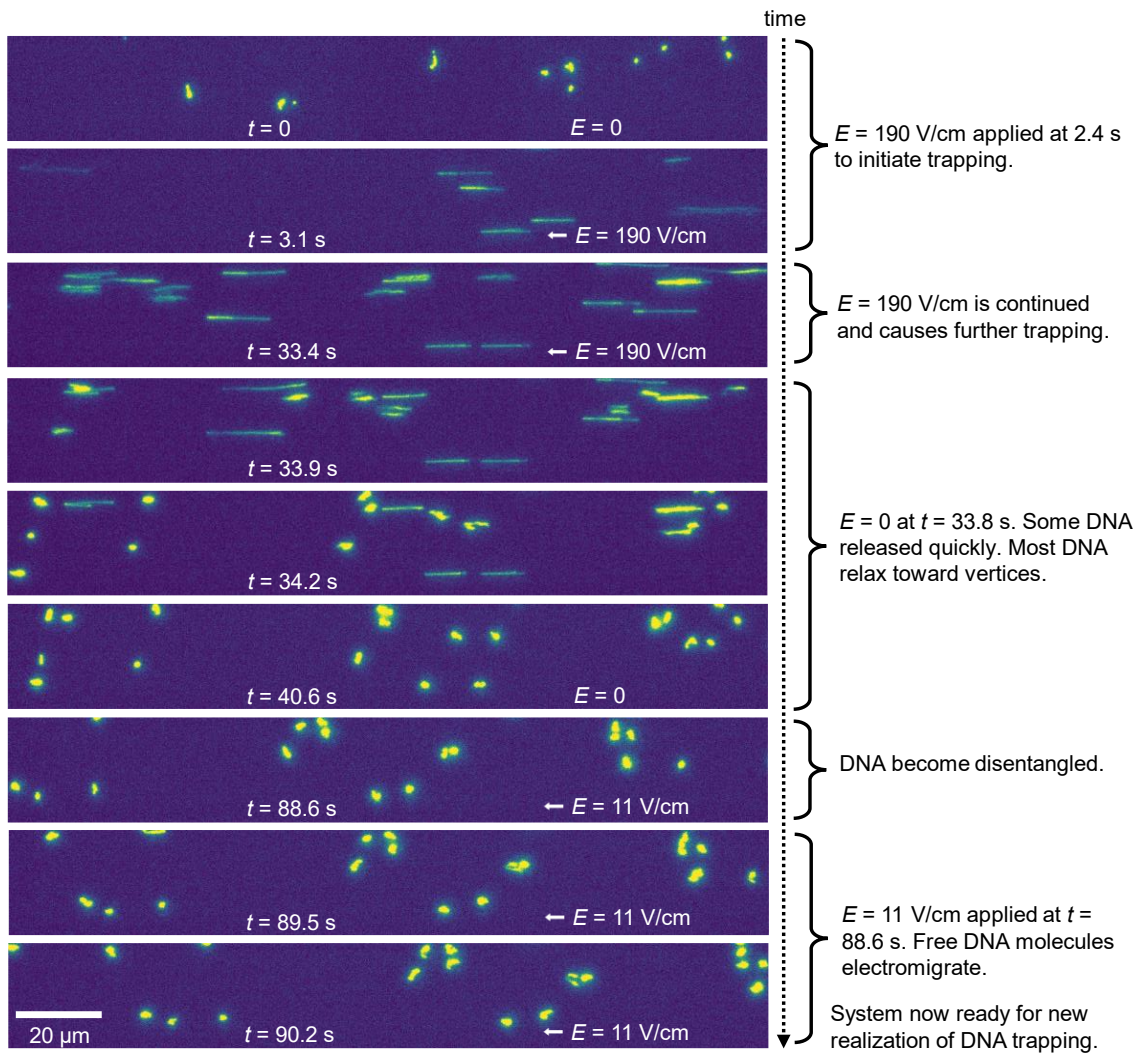

**Extended Data Fig. 6 | Representative cycle of DNA vertex-pinning, relaxation, disentanglement, and free electromigration.** Consecutive (top to bottom) images of single-molecule DNA. DNA molecules are initially in expected random coil shapes. $E$ = 190 V/cm was applied at $t$ = 2.4 s to initiate vertex pinning. $E$ deactivated at $t$ = 33.8 s. Most DNA relaxed toward vertices. After about 50 s, more than 95% of DNA became unpinned from the wall. At $t$ = 88.6 s, a relatively low axial electric field of 11 V/cm was applied to electromigrate away previously vertex-pinned DNA molecules. Note the displacements of correlated patterns across the last three images.

## Supporting Information for

Vertex pinning and stretching of single molecule DNA in a linear polymer solution

Kunlin Ma[a], Caleb J. Samuel[a], Soumyadeep Paul[a], Fereshteh L. Memarian[b], Gabrielle Vukasin[b], Armin Darvish[b], Juan G. Santiago[a]*

[a] Department of Mechanical Engineering, Stanford University, Palo Alto, CA 94305.

[b] Robert Bosch LLC, Research & Technology Center, Sunnyvale, CA 94085.

* Corresponding author: Juan G. Santiago

**Email:**  juans@stanford.edu

**This file includes:**

- Supporting text
- Figures S1 to S6
- Legends for Movies S1 to S10
- SI References

**Other supporting materials for this manuscript include the following:**

- Movies S1 to S10

**Supporting Information Text**

**S1 Methods**
**S1.1    Image processing for quantification of amount of vertex pinned DNA**
   All data in the image data in this section was acquired using the aforementioned water immersion 60× objective (OLYMPUS UPlanApo) with a numerical aperture (NA) of 1.2. The camera here was Hamamatsu Orca-Flash 4.0LT sCMOS camera. This camera has a sensor pixel size of 6.5 by 6.5 µm. The pixel size in image plane was approximately 0.11 by 0.11 µm per pixel. The camera acquired 20 frame per second (fps) with an exposure time of 49 ms. Each realization's raw TIFF sequences were imported into MATLAB for image analysis. Each 30 s sequence consisted of 600 images. First, we applied a moving (temporal) median operation. This moving median replaces the pixel value of each image with the median value of the same pixel across a small number of previous and subsequent images. This operation has the effect of enhancing and retaining approximately stationary image data (pinned, stretched DNA) and rejecting moving objects (electromigrating DNA). The highest applied electric fields of 363 V/cm were associated with the highest DNA velocities. Hence, each median operation considered only a three-image group (i.e., each image pixel value was replaced by the median among the previous frame, current frame, and preceding frame). The median operation for lower applied electric fields (slower moving DNA) used a progressively larger number of images within each group. We considered a maximum of 21 frames for the lowest applied fields of 19 V/cm. The top subplot of Figure 3A (of the main paper) shows an example frame obtained from the moving median image sequence. See especially Movie S8 for a comparison between raw video versus moving median video, and note how the image data from stationary objects (pinned DNA) are enhanced. Next, we used Bradley's method to perform local mean-based adaptive thresholding on the moving median image sequences[31]. To this end, we used the MATLAB command *adaptthresh* with a sensitivity parameter of 0.45 followed by *imbinarize* to obtain binarized images for each frame. The pixel values of the binarized images were set to zero below the threshold and to unity above the threshold. Next, we identified and stored the data for the thresholded pixel coordinates via the MATLAB command *find*. We then performed an alpha shape operation on the binarized image in order to efficiently reject non-DNA image noise from the analysis. Alpha shape is a generalization of the concept of convex hull identification and is a subset of Delaunay triangulation [16]. The alpha shape method is widely used for pattern recognition in computational graphics to identify the relatedness (via proximity) of points on a graph[32]. To obtain the alpha shapes, we used the binarized pixel coordinates and an alpha radius of 2.5 as inputs to the MATLAB command *alphaShape*. We then projected the resulting alpha shapes onto a binary mask. The binary mask was first dilated and then eroded with the same 10-pixel long streamwise line as a structuring element (again to favor even field-aligned and pinned DNA images). After erosion, our algorithm obtained alpha shape bounding masks. The alpha shape method successfully reduced background noise caused by adaptive thresholding. The alpha shape operation and the dilation and erosion also smoothed the noisy edges of the identified objects. The middle sub-image of Figure 3A (of the main paper) shows an example of the resulting mask. At each time frame, the binarized mask was used on the moving median image to obtain a sequence of thresholded moving median images. The bottom sub-image of Figure 3A shows an example of the resulting operation between the alpha-shape-bounded mask and one of the median images. The (original, unchanged) intensity of detected DNA within the resulting thresholded images was then integrated spatially to acquire a scalar signal which varied with time at the frame rate of 20 fps. We term this scalar $\langle I \rangle$ in Figure 3 of the main paper and in the discussions and figures below.

**S1.2    Image processing for analysis of relaxation of vertex-pinned DNA**
   This section describes the image processing analyses performed to obtain the DNA relaxation data shown in Figures 4B, 4C and 4D of the main text. First, TIFF sequences were gaussian filtered temporally with a three-frame filter using MATLAB built-in function *imgaussfilt3*.

Then, we followed the same method explained above to obtain alpha shape bounding masks for videos of DNA relaxation, except we did not include dilation followed by erosion in order to best preserve DNA shape. Once we obtained the bounding masks, we used the MATLAB command *bwconncomp* followed by *regionprops* to determine morphological and location properties of distinct connected regions inside the alpha shape bounding mask. In an effort to remove image noise and very small DNA fragments from remaining in the following analyses, we removed connected regions with total intensity less than 100,000 and 20,000 for 48.5 and 20 kbp DNA sample, respectively. These thresholds were designed to remove objects with significantly less intensity than largely intact pieces of DNA. To remove image noise and very small DNA fragments from remaining in the following analyses, we removed connected regions with integrated intensity values which were lower than 2.5 standard deviations from the mean. Figures S4A to S4C and Figures S4F to S4H show examples of the aforementioned image processing procedures for vertex-pinned DNA before and during relaxation. Then, again using the MATLAB command *regionprops*, we fitted an ellipse to each remaining region. These ellipses had the same normalized second central moments as the regions to which they were fit. Examples of the ellipses fitted to various pieces of DNA can be seen in Figure 4B. For each realization of DNA relaxation at each frame after electric field was turned off, we then had a collection of fitted ellipses ideally representing the shape of various pieces of DNA at that given time for that given realization. Figures S4D to S4E and S4I to S4J show examples of the aforementioned island identification and major axes estimation. Figures S4D and S4E show an example for vertex-pinned DNA before relaxation while Figures S4I and S4J show an example for vertex-pinned DNA during relaxation. The ellipses approximation is shown in red curves, and the major axes are shown in green lines. From this collection of fitted ellipses, we recorded the following two values: the mean length of the fitted ellipses' major axis, which we used as the characteristic length of DNA during relaxation; and the standard deviation of major axis length for fitted ellipses. The resulting analyses are shown in Figure 4C and 4D for one and two standard deviations. The black line shows the characteristic length, and the blue and green shaded regions represent data with one or two standard deviations of the mean major axis length. The only difference between the relaxation analyses of 48.5 and 20 kbp DNA was the threshold chosen to remove selected regions we believed to be DNA fragments or image noise. Figures S5A to S5J shows the same image processing for 20 kbp DNA data as described above of Figure S4A to S4J for 48.5 kbp DNA data.

In addition to Figure 4, we include two raw epifluorescence videos of 48.5 and 20 kbp DNA relaxing (Movie S3 and S4) and another processed video of approximately 300 superposed images of DNA relaxing for 48.5 kbp DNA (see Movie S9). To generate superposed images of DNA relaxing for Movie S9, we first obtained alpha shape bounding masks as described previously after the electric field was turned off. Next, we used MATLAB command *regionprops* to remove regions with less than 50,000 total integrated intensity. After this, we used the MATLAB command *bwlabel* to identify and label distinct connected regions inside the alpha shape bounding mask. For each region, we then used *find* command to generate a list of pixels within that region. From this list, we estimated the most "upstream" pixel as the vertex-pinned point, as well as the median pixel location in the spanwise direction. We then applied the alpha shape bounding mask to the original image so that each previously identified region had its original intensity values. We then added the original image intensities within these alpha shape regions for each time in the relaxation images.

**S1.3   Colocalization analyses of multiple realizations of DNA vertex-pinning**

To further analyze the repeatability of DNA pinning sites, we performed colocalization analysis based on the data from Figs. 5A, 5B and 5C in the main text. Fig. 6 shows the resultant colocalization scatter plots and the respective computed Pearson correlation coefficients. To create these plots, we first read two raw TIFF image data (for example Figs. 5A and 5B) into arrays in MATLAB. These images were registered such that individual pixel locations were the same across realizations. For each pixel, we store image data in a two-dimensional intensity histogram where the horizontal axis value corresponds to the intensity value from the first realization image(abscissa) and the vertical axis value corresponds to the other second realization image (ordinate). The 2D histograms obtained thusly are shown in false color in Figure 5A, 5B and 5C for each pair of raw images. The Pearson correlation coefficients (PCC) are calculated based on the correlation data[30]. The PCC values are respectively 0.30, 0.25 and 0.22 for the three plots shown. These correlation values and the shape of the correlation diagram indicate that the data have very poor colocalization of features. The lowest intensities (bottom left region) are correlated across realizations because they represent the background signal (image regions where there is no DNA). The low colocalization counts at the high intensity range suggest poor correlation of DNA trapping sites between realizations.

As a comparison case, we performed the same colocalization analysis for three images taken from a single realization. Figure S6A shows three raw epi-fluorescence images from one realization of DNA vertex-pinning under high electric field. The first and the second sub-panels are 5 s apart and the second, and the third sub-panel is 2 s after the second. Figures S6B and S6C show the colocalization scatter plot comparing DNA trapping at $t_1$ = 21 s to $t_2$ = 26 s, and at $t_2$ = 26 s to $t_3$ = 28 s. Note the strong correlation along the diagonal of these plots and compare these to Fig. 5 and Fig. 6 in the main manuscript. The computed PCC values here are 0.81 and 0.86, respectively, indicating a high degree of spatial correlation. These comparisons are further evidence that the trapping sites for DNA vertex pinning are uncorrelated across different realizations.

### S1.4 Spatial averaging intensity

In the main text and SI, the symbol $\langle \cdots \rangle$ is used to indicate a spatial averaging over all pixels within a two-dimensional region. For example, to indicate spatial integration of intensity values of a raw image or two-dimensional correlation. For a quantity $q$ for each time frame, the spatial averaged value is then denoted as

$$\langle q \rangle = \frac{1}{N_x N_y} \sum_{i=1}^{N_x} \sum_{j=1}^{N_y} q_{ij}, \tag{1}$$

where $N_x$ and $N_y$ indicate the pixel number in the horizontal and vertical directions, respectively.

### S1.5 Estimation of the magnitude of the axial electric field

The electric field measurements reported in this work were determined from ionic current measurements and measurements of the local channel cross section. We measured the conductivity of the buffer as described previously using a commercial conductivity meter (Orion VersaStar Pro Advanced Electrochemistry Meter, ThermoScientifc). We measured the current during the experiments via a Keithley 2400 SourceMeter. The transverse depth of the channel was measured via a profilometer as 0.9 µm. The channel spanwise width was determined from bright field images as we show in as the left most subplot in Figure 1C. This width was 20 µm. Assuming ionic current is dominant, the axial electric field strength is related to the current through Ohmic law:

$$j = \sigma E, \tag{2}$$

where $j$ is the current density and $\sigma$ is the conductivity of the buffer. We divided the measured current, $I$, by the estimated cross-sectional area, $A$, to obtain the current density, $j$. Then, we estimate the electric field strength by $E = j/\sigma$.

**S2. Materials**

**S2.1    Contour length of PVP and DNA of multiple MWs**

We note that unstained $\lambda$-DNA (48.5 kbp) and 20 kbp DNA have reported contour lengths of approximately 16 and 6.6 µm, respectively. When $\lambda$-DNA and 20 kbp DNA are stained with YOYO-1 dye, their full contour length is extended to be approximately 21 and 8.7 µm, respectively[33]. PVP of MWs of 10, 58, 360, and 1300 kDa have contour lengths of approximately 25, 145, 900, and 3250 nm, respectively[34,35].

**S2.2    Range of DNA sizes for vertex-pinning**

We have used the aforementioned method to trap DNA of sizes of 48.5, 20, 10, and 1 kbp under sufficiently high electric fields (data not shown). However, we do not know the minimum size of DNA that can be vertex pinned since we cannot acquire clear visualization of smaller sizes of DNA.

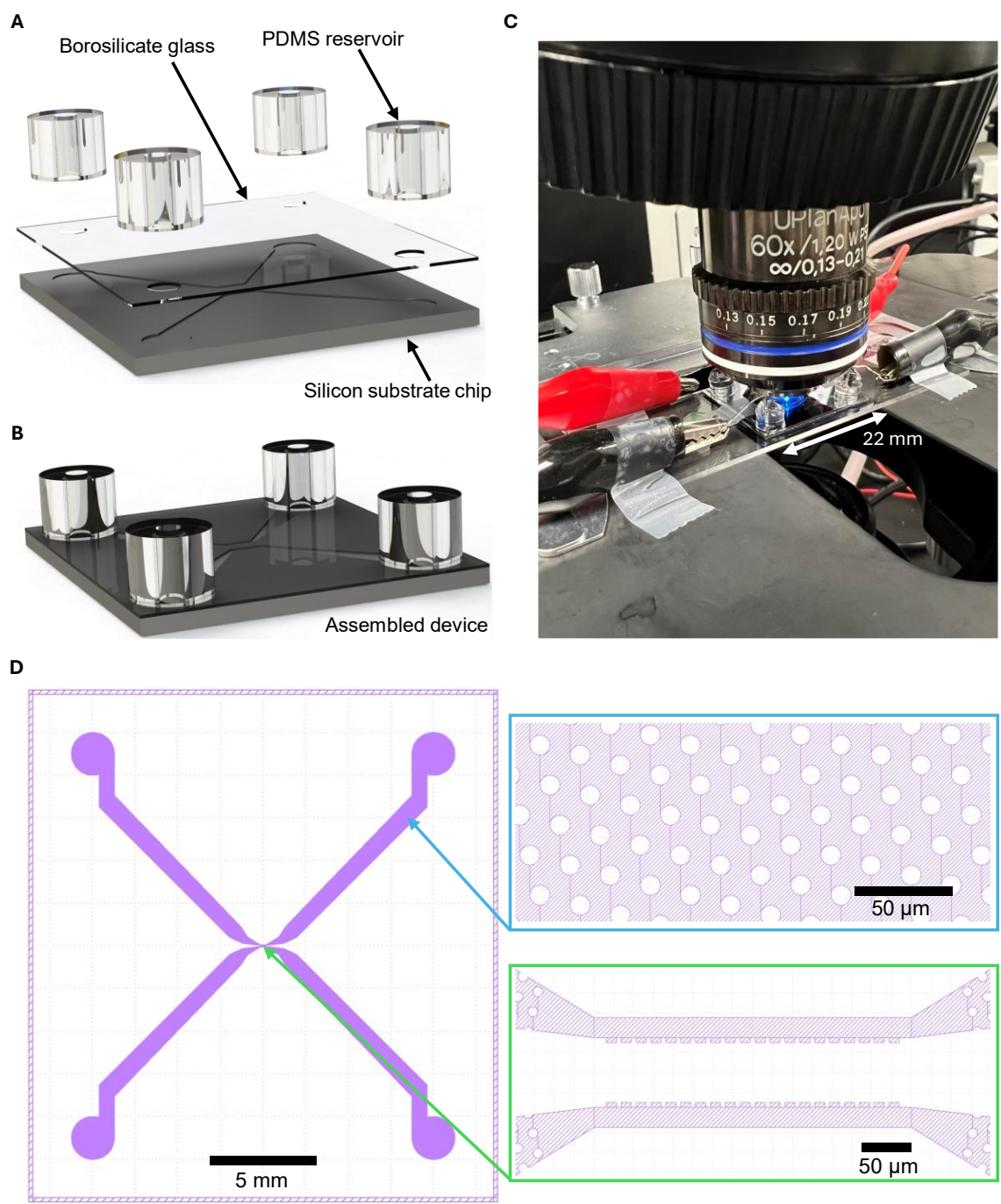

**Fig. S1 | Custom microfluidic device and apparatus.** (**A**) Pieces of PDMS reservoirs, 170 µm thick borosilicate glass and silicon substrate chips are shown as the three major components of the custom microfluidic device. (**B**) Assembled custom microfluidic device. Borosilicate glass was anodically bonded to the silicon substrate chip and then PDMS reservoirs were plasma bonded to the borosilicate glass. (**C**) The assembled custom microfluidic device was taped to a glass slide and then visualized via a water immersion 60× objective of numerical aperture of 1.2. The illumination source is a blue LED. Platinum electrodes were inserted into the reservoir filled with liquid. (**D**) The layout of whole silicon substrate chips and zoomed in images of two regions are

shown. The center region has castellation features near one side. The large microchannel regions leading to reservoirs have posts arrays of diameter of 10 µm. The posts array is designed to support the anodically bonded borosilicate glass.

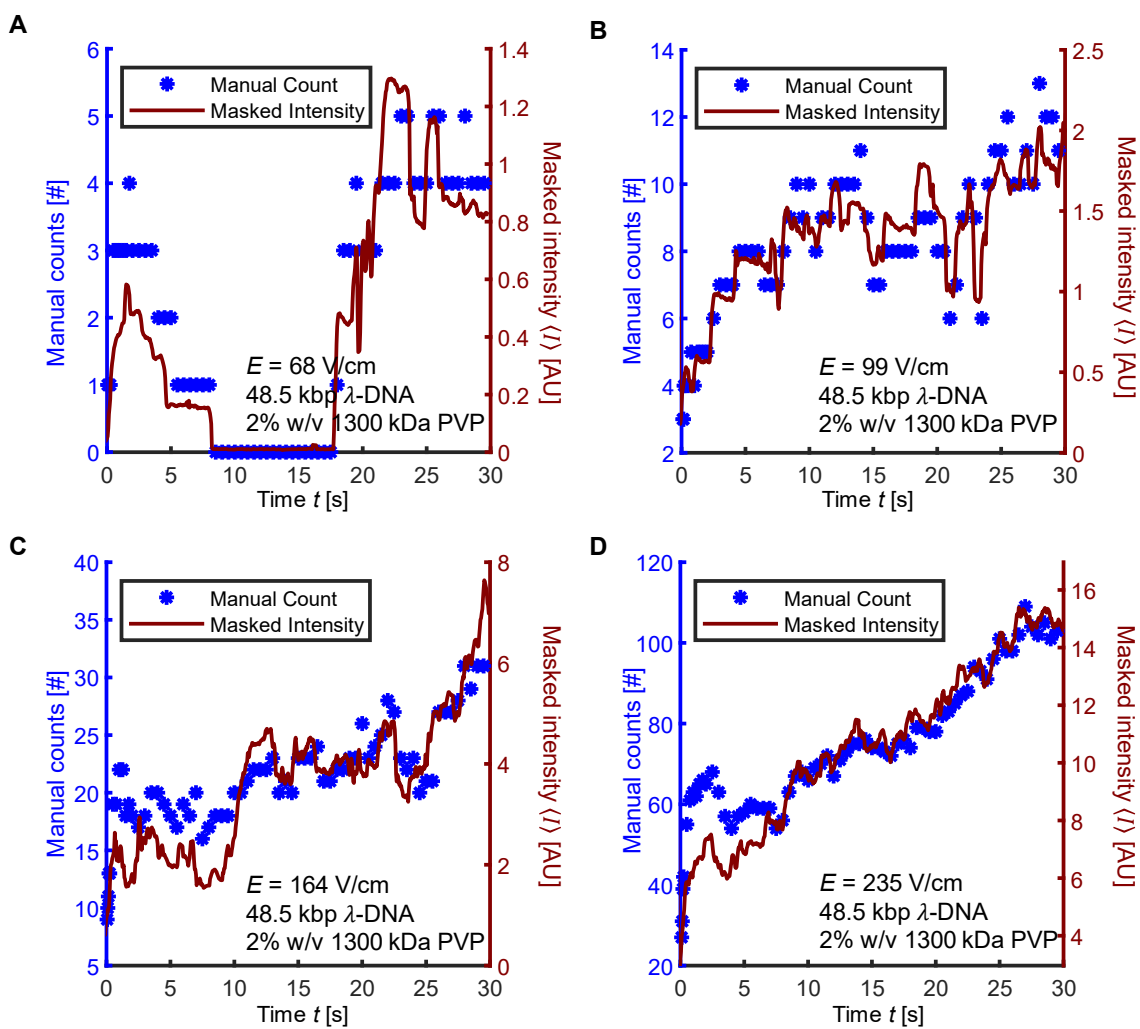

**Fig. S2 | Comparison between area-averaged alpha-shape-bounded intensity ⟨*I*⟩ versus results from manual counting of molecules.** (**A** to **D**) Red curves show the alpha shape bounding intensity versus time for electric field strength of 18, 99, 164, and 235 V/cm, respectively. Blue scatter points show the manual counting of each entangled DNA at the corresponding time step for electric field strength of 18, 99, 164 and 235 V/cm, respectively. All data are 4 pM of YOYO-1 labeled λ-DNA (48.5 kbp) in 1×TBE buffer with 4% v/v β-mercaptoethanol added with 1300 kDa PVP polymer. The alpha shape bounding intensity versus time have trends that are similar to the manual counting results.

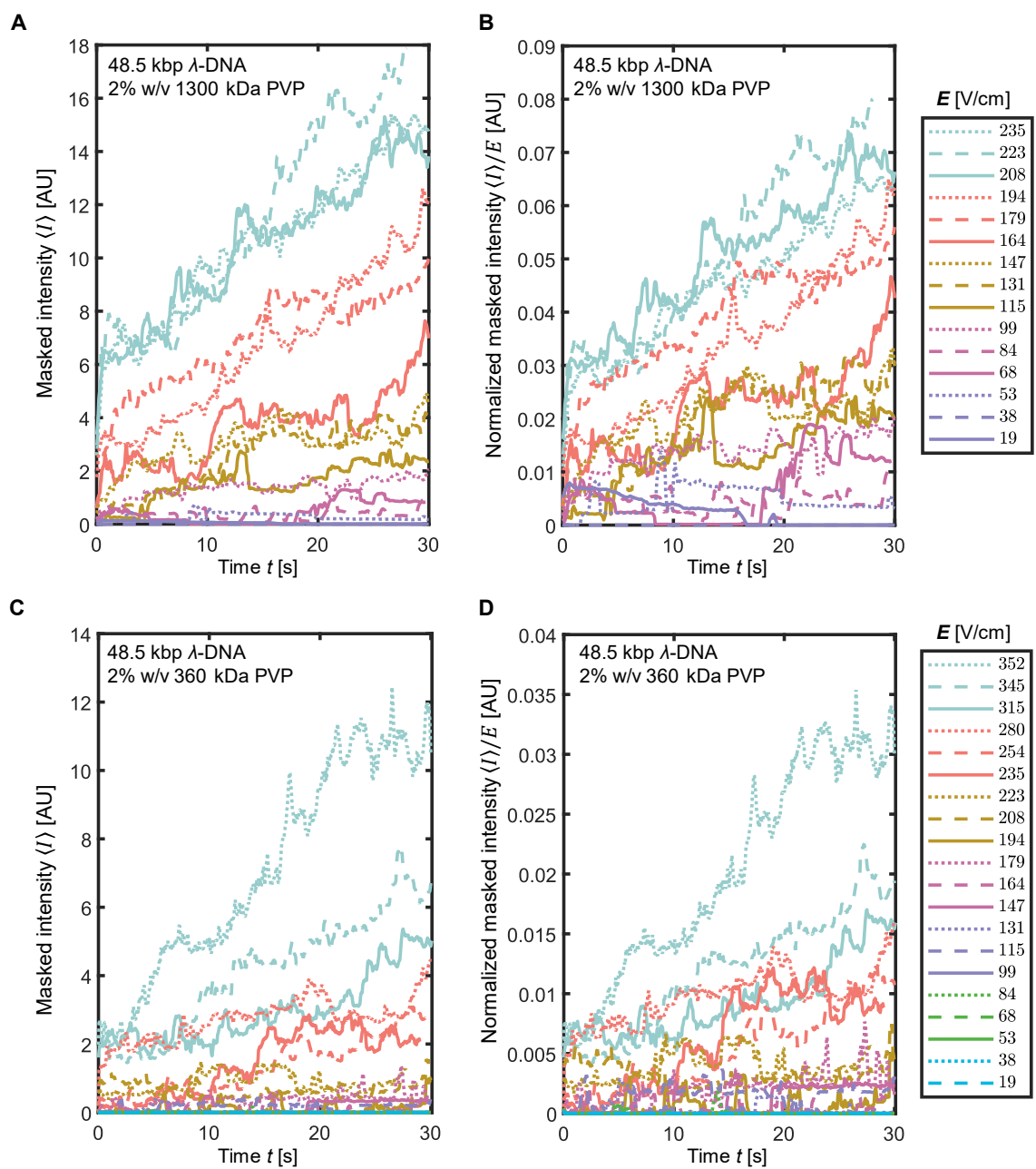

**Fig. S3 | Additional experimental quantification of the amount of vertex-pinned, single-molecule DNA as a function of time and field strength.** (**A** to **D**) Alpha shape bounding masked intensity versus time are shown for realizations performed at multiple respective electric fields. All data are 4 pM of YOYO-1 labeled λ-DNA (48.5 kbp) in 1×TBE buffer with 4% v/v β-mercaptoethanol. (**A** and **B**) have an addition of 2% w/w 1300 kDa PVP polymer in the buffer and (**C** and **D**) have an addition of 2% w/w 360 kDa PVP polymer in the buffer. Intensity in (**B** and **D**) are normalized by the applied axial electric field strength for each realization.
33

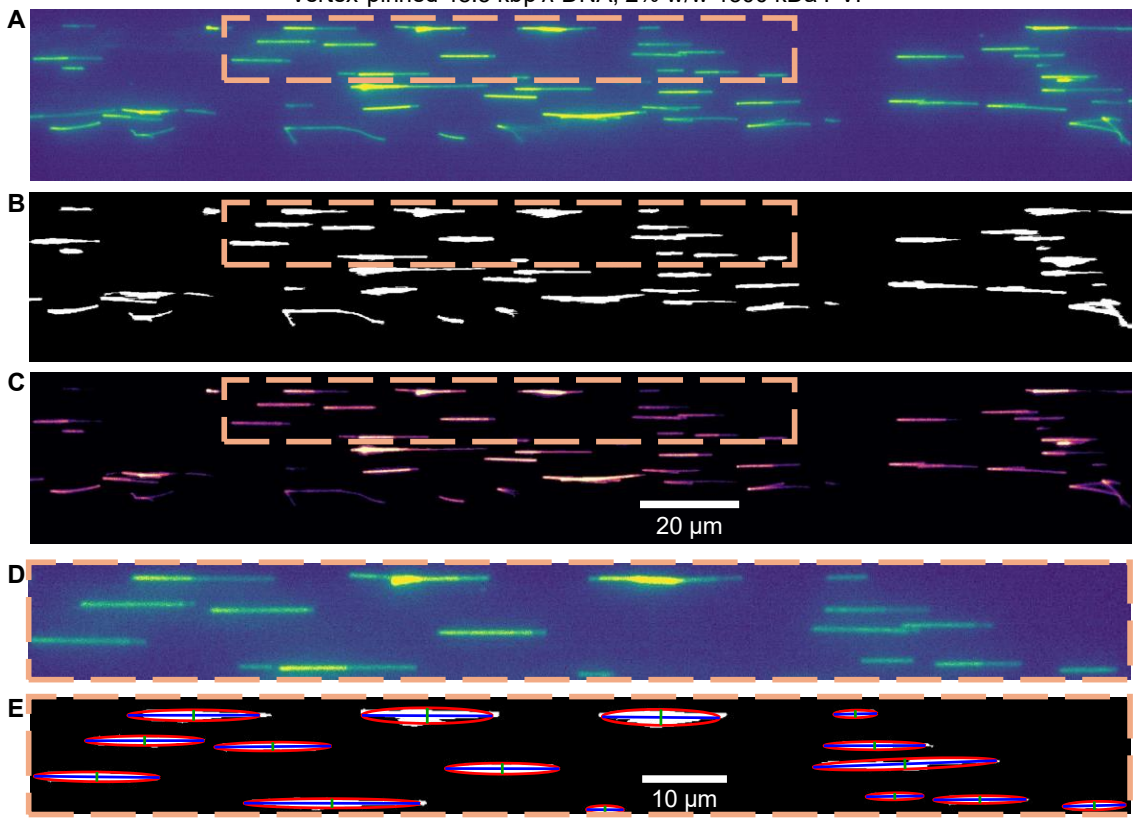

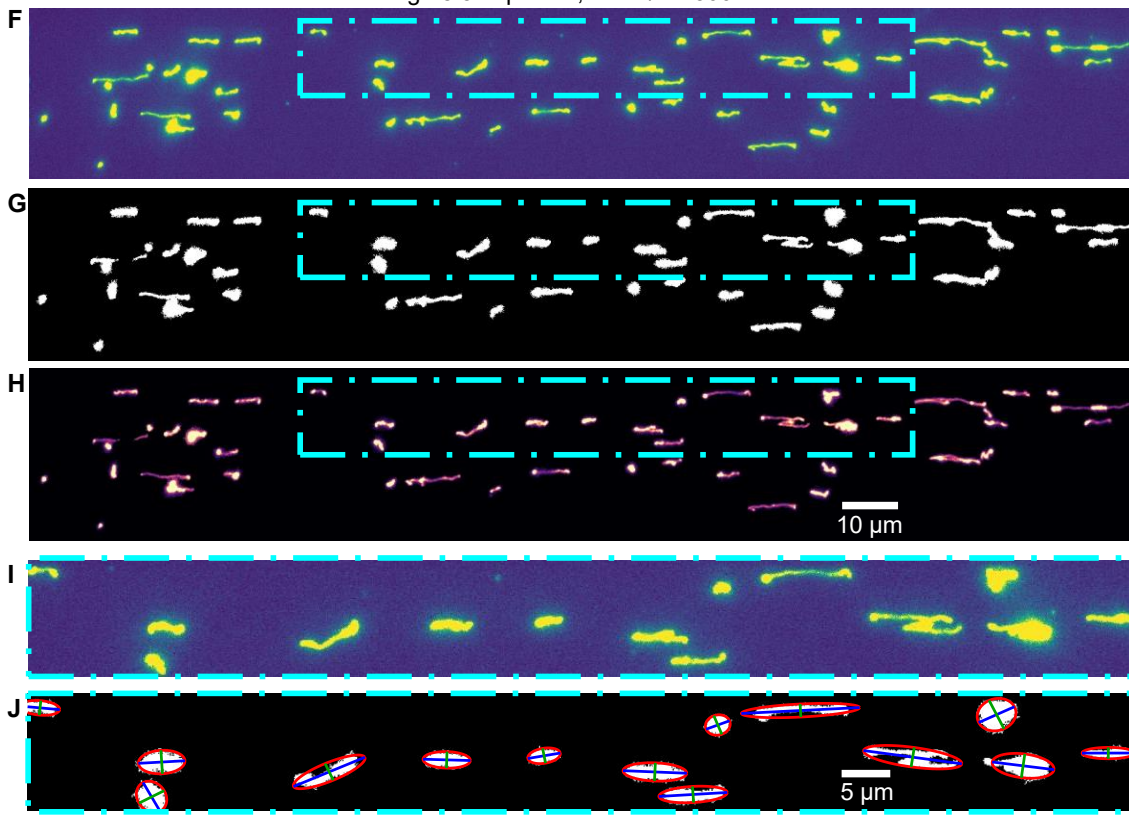

**Fig. S4 | Summary of automated image processing used to quantify 48.5 kbp DNA relaxation.** (**A**) Example image of vertex-pinned DNA before relaxation. (**B**) Binarized image of (A) after adaptive thresholding and alpha shape operation. (**C**) Image was obtained by integrating (A) with the mask in (B). (**D**) Zoomed in image of (A). (**E**) Example image to show the auto detection of each island. The red curve shows the result of ellipse shape approximation. The blue lines and the green lines show the major and secondary axis of each approximated ellipse shape, respectively. (**F** to **J**) Images showing the same procedure as from (A) to (E), except for the image was chosen during the DNA relaxation process.

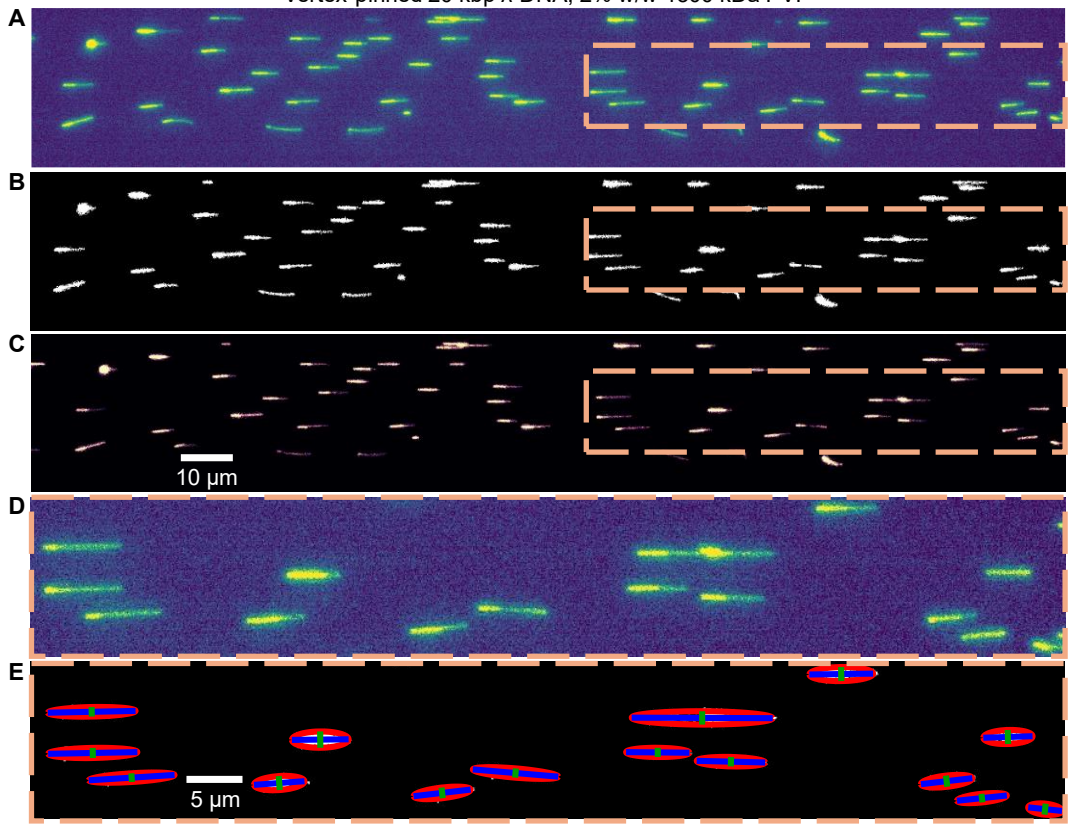

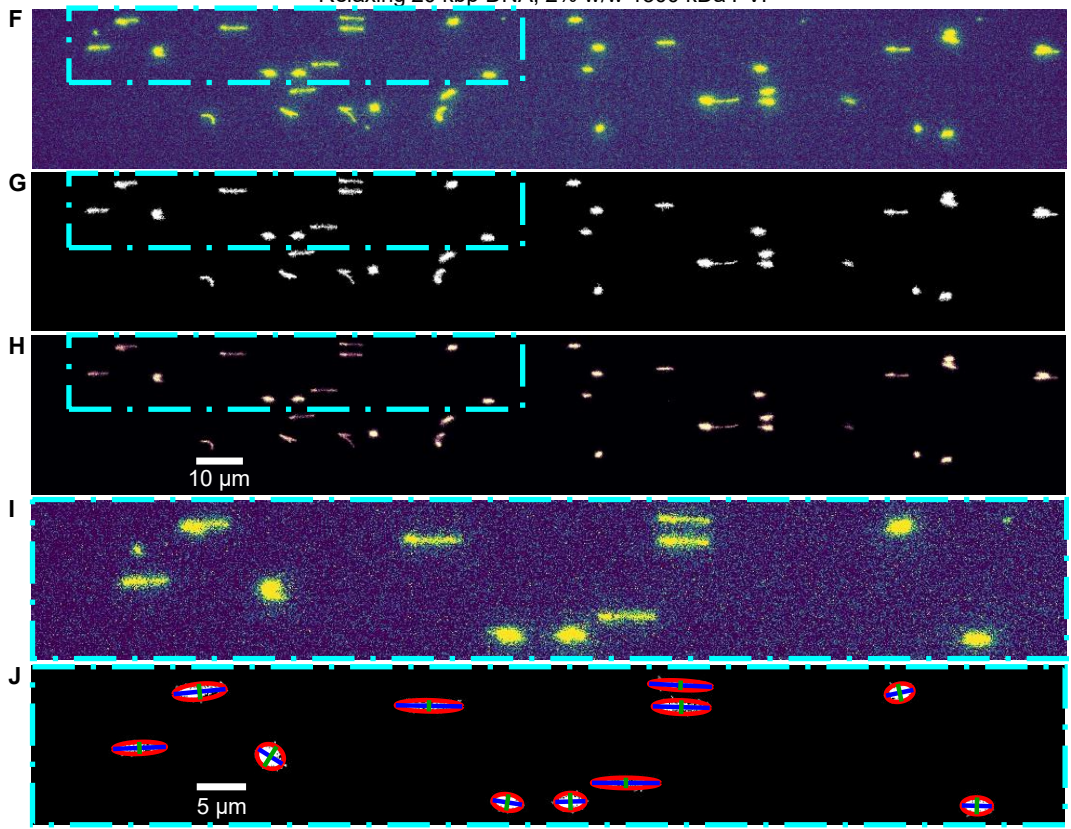

**Fig. S5 | Summary of automated image processing used to quantify 20 kbp DNA relaxation.**
(**A**) Example image of vertex-pinned DNA before relaxation. (**B**) Binarized image of (A) after adaptive thresholding and alpha shape operation. (**C**) Image was obtained by integrating (A) with the mask in (B). (**D**) Zoomed in image of (A). (**E**) Example image to show the auto detection of each island. The red curve shows the result of ellipse shape approximation. The blue lines and the green lines show the major and secondary axis of each approximated ellipse shape, respectively. (**F** to **J**) Images showing the same procedure as from (A) to (E), except for the image was chosen during the DNA relaxation process.

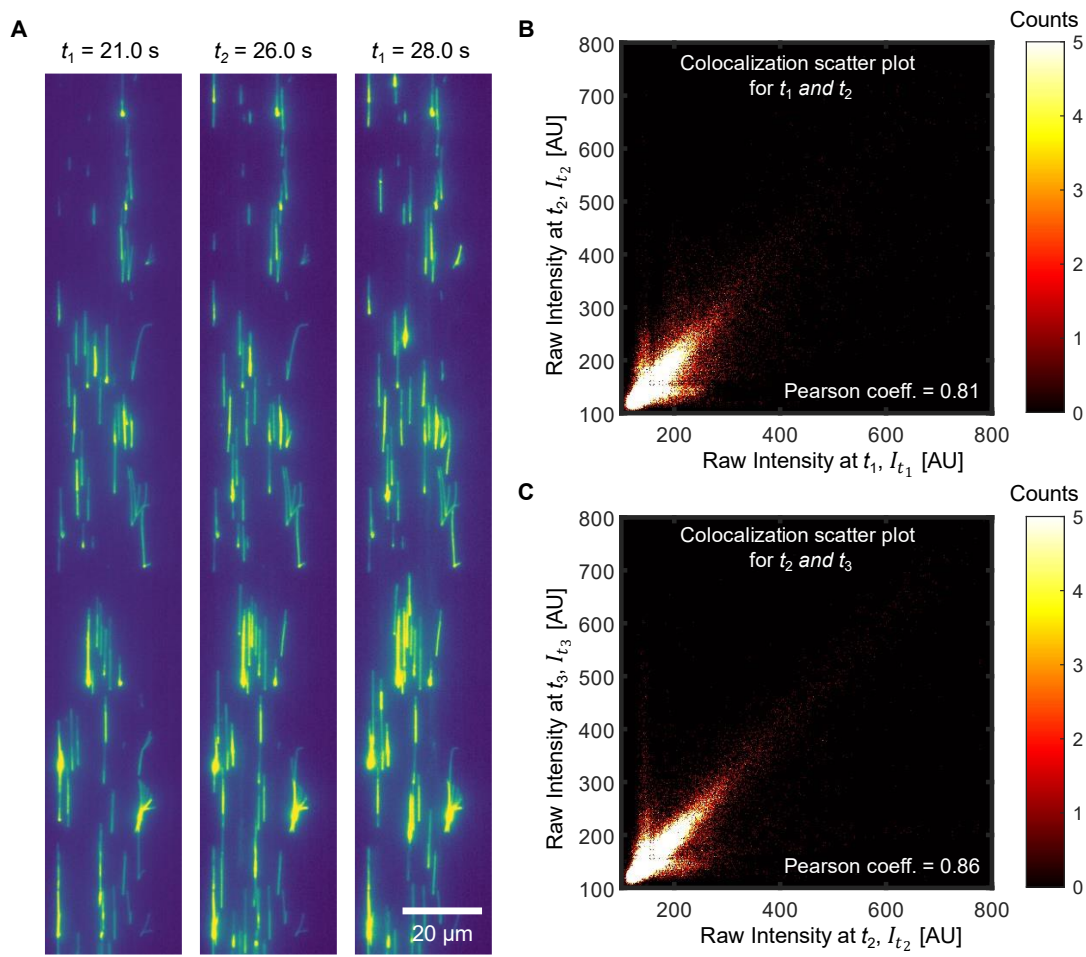

**Fig. S6 | Colocalization analysis of image intensity of vertex-pinned DNA across three consecutive images from one realization.** (**A**) Three raw consecutive epi-fluorescence images are shown for a single realization of DNA vertex-pinning experiments. The first sub-panel and the second sub-panel are 5 s apart. The second and third sub-panels are 2 s apart. (**B** to **C**) The colocalization scatter plots (correlation diagrams) compare each consecutive image pair. The computed Pearson correlation coefficients values for the two time-sequenced image pairs are 0.81 and 0.86, respectively.

**Movie S1 (separate file).**

Video was acquired with fluorescently labeled $\lambda$-DNA (48.5 kbp) migration in 1×TBE buffer with 4% v/v $\beta$-mercaptoethanol and addition of 2% w/w 1300 kDa PVP. The DNA molecules were subjected to an axial electric field of 68.5 V/cm from right to left, which was the approximate electric field strength suggested by Figure 3. The channel had a geometry as shown in Figure S1D and had a depth of 0.9 μm. The raw video was shown on the top part of the video. The moving median video was shown on the bottom part of the video and was processed as described in the section "Image processing for quantification of amount of vertex pinned DNA". Note how the intensity data from stationary objects (pinned DNA) are enhanced while the intensity from moving objects were suppressed.

**Movie S2 (separate file).**

Video was acquired with fluorescently labeled $\lambda$-DNA (48.5 kbp) migration in 1×TBE buffer with 4% v/v $\beta$-mercaptoethanol and addition of 2% w/w 360 kDa PVP. The DNA molecules were subjected to an axial electric field of 179 V/cm from right to left, which was the approximate electric field strength suggested by Figure 3. The channel had a geometry as shown in Figure S1D and had a depth of 0.9 μm. The raw video was shown on the top part of the video. The moving median video was shown on the bottom part of the video and was processed as described in the section "Image processing for quantification of amount of vertex pinned DNA". Note how the intensity data from stationary objects (pinned DNA) are enhanced while the intensity from moving objects were suppressed.

**Movie S3 (separate file).**

Video was acquired with fluorescently labeled $\lambda$-DNA (48.5 kbp) migration in 1×TBE buffer with 4% v/v $\beta$-mercaptoethanol and addition of 2% w/w 1300 kDa PVP. An electric field of 194 V/cm was applied from $t$ = 0 to 9.6 s to allow DNA to be vertex-pinned and the electric field was turned off afterward to allow DNA to relax and disentangle. Some of the vertex-pinned DNA relaxed toward the trapping point, and some of the vertex-pinned DNA was quickly released from the trapping point. Note after switching off the electric field, some cloud shape DNA molecules entered the field of view by electromigration without trapping.

**Movie S4 (separate file).**

Video was acquired with fluorescently labeled 20 kbp DNA migration in 1×TBE buffer with 4% v/v $\beta$-mercaptoethanol and addition of 2% w/w 1300 kDa PVP. An electric field of 235 V/cm was applied from $t$ = 0 to 9.6 s to allow DNA to be vertex-pinned and the electric field was turned off afterward to allow DNA to relax and disentangle. Some of the vertex-pinned DNA relaxed toward the trapping point, and some of the vertex-pinned DNA was quickly released from the trapping point. Note after switching off the electric field, some cloud shape DNA molecules entered the field of view by electromigration without trapping.

**Movie S5 (separate file).**

Video was acquired with fluorescently labeled $\lambda$-DNA (48.5 kbp) migration in 1×TBE buffer with 4% v/v $\beta$-mercaptoethanol and addition of 2% w/w 1300 kDa PVP. Two isolated channels of geometry as shown in Figure S1D were imaged and both had a depth of 0.9 μm. Samples in both branches were the same, but DNA in top branch was driven by pressure driven flow only and the DNA in bottom branch was driven by electric fields only. The applied electric field at the bottom branch was initially 410 V/cm for DNA vertex-pinning and then canceled to zero to allow DNA to relax and disentangle over around 50 s. Next, 12 V/cm of electric field was applied to clear out the previously trapped DNA molecules and introduced new ones.

**Movie S6 (separate file).**

Video was acquired with fluorescently labeled $\lambda$-DNA (48.5 kbp) migration in 1×TBE buffer with 4% v/v $\beta$-mercaptoethanol and addition of 2% w/w 1300 kDa PVP. The DNA molecules were subjected to an axial electric field suddenly increased from 10 V/cm to 140 V/cm. The channel was an approximately 2.2 cm long straight channel with a cross-section of 40 µm wide and 3 µm deep. The depth of field was estimated to be 0.7 µm. The focal plane was adjusted during the experiment to show DNA vertex-pinning at top and bottom surfaces.

**Movie S7 (separate file).**

Video was acquired with fluorescently labeled $\lambda$-DNA (48.5 kbp) migration in 1×TBE buffer with 4% v/v $\beta$-mercaptoethanol and addition of 2% w/w 1300 kDa PVP. Two isolated channels of geometry as shown in Figure S1D were imaged and both had a depth of 0.9 µm. Samples in both branches were the same, but DNA in top branch was driven by pressure driven flow only and the DNA in bottom branch was driven by electric fields only. The applied electric field at the bottom branch was switching between 190 V/cm and 0 V/cm to demonstrate DNA vertex-pinning at sufficiently high electric field comparing to no vertex-pinning under pressure driven flow.

**Movie S8 (separate file).**

Video was acquired with fluorescently labeled $\lambda$-DNA (48.5 kbp) migration in 1×TBE buffer with 4% v/v $\beta$-mercaptoethanol and addition of 2% w/w 1300 kDa PVP. The DNA molecules were subjected to an axial electric field of 147 V/cm from right to left. The channel had a geometry as shown in Figure S1D and had a depth of 0.9 µm. The raw video was shown on the top part of the video. The moving median video was shown on the bottom part of the video, and was processed as described in the section "Image processing for quantification of amount of vertex pinned DNA". Note how the intensity data from stationary objects (pinned DNA) are enhanced while the intensity from moving objects were suppressed.

**Movie S9 (separate file).**

Superposing of approximately 300 DNA during relaxation after being vertex-pinned. The trapping point (vertex) was superposed, and the direction of DNA molecules were aligned. The detail image processing procedure was discussed in the SI. The sample used is fluorescently labeled $\lambda$-DNA (48.5 kbp) in 1×TBE buffer with 4% v/v $\beta$-mercaptoethanol and addition of 2% w/w 1300 kDa PVP.

**Movie S10 (separate file).**

Video was acquired with fluorescently labeled $\lambda$-DNA (48.5 kbp) migration in 1×TBE buffer with 4% v/v $\beta$-mercaptoethanol and addition of 2% w/w 1300 kDa PVP. The DNA molecules were subjected to an axial electric field of 256 V/cm from right to left. The channel had a geometry as described by Extended Data Fig. 2 and had a width of 100 µm and a depth of 37 µm, approximately. The channel was all made by glass. Clouds of DNA molecules were electromigrated while some DNA molecules were vertex pinned in the background.

**SI References**